\documentclass[12pt,colorinlistoftodos]{article}
\usepackage{amsmath}
\usepackage{amsthm}
\usepackage{graphicx} 
\usepackage{color}
\usepackage{todonotes}
\usepackage{float}
\newtheorem{theorem}{Theorem}
\newtheorem{lemma}[theorem]{Lemma}

\usepackage{subcaption}    

\usepackage{todonotes}

\def\IR{\mathbb{R}}
\def\bE{\bm{E}}
\def\bF{\mathbf{F}}

\def\bz{\bm{z}}
\def\b0{\bm{0}}

\def\R{\mathcal{R}}

\usepackage{booktabs}
\usepackage{hyperref}
\usepackage{bm}
\usepackage{doi}

\hypersetup{
    colorlinks=true,
    linkcolor=blue,
    citecolor=blue,
    urlcolor=blue
}
\usepackage{tcolorbox}

\usepackage{amssymb}
\title{Threshold-based impulsive biocontrol for coffee leaf rust}
\author{Clotilde Djuikem \and Julien Arino}
\date{Version of \today}

\usepackage{amsmath}
\usepackage{enumitem}
\usepackage{geometry}
\begin{document}
\maketitle

\begin{abstract}  
Coffee leaf rust (CLR) severely affects coffee production worldwide, leading to reduced yields and economic losses. 
To reduce the cost of control, small-scale farmers often only apply control measures once a noticeable level of infection is reached.
In this work, we  develop  mathematical models to better understand CLR dynamics and impulsive biocontrol with threshold-based interventions.
We first use ordinary and impulsive differential equations to describe disease spread and the application of control measures once a certain infection level is detected. 
These models help determine when and how often interventions should occur.
To capture the early stages of the disease and the chance that it might die out by itself, we then use a continuous-time Markov chain  approach. 
This stochastic model allows us to estimate the probability that the pathogen fails to establish, thereby avoiding serious outbreaks.
\end{abstract}

\section{Introduction}
\label{sec:introduction}
Coffee leaf rust (CLR), caused by the fungal pathogen \textit{Hemileia vastatrix}, is among the most devastating diseases affecting coffee production worldwide \cite{avelino2015impact, mccook2019history}. 
First recorded in Sri Lanka in the late 19th century, CLR rapidly spread across coffee-growing regions due to its adaptability and high dispersal capacity \cite{mccook2019history}. 
The disease causes severe defoliation, reducing coffee plant productivity by diminishing photosynthesis and leads to yield losses of up to 70\% in heavily infected plantations and particularly for smallholder farmers \cite{DABA2022e11892, DUPRE2022105918}. 
This impact is particularly acute in smallholder farming systems, which constitute the backbone of global coffee production \cite{jaramillo2013climateimpact}. 
Despite efforts in disease management, CLR remains challenging to control due to its rapid spread and adaptability. 

Traditional control measures include cultural practices such as pruning, maintaining optimal shade levels and planting resistant varieties \cite{zambolim2016current}. 
However, these methods are often insufficient when faced with large-scale outbreaks, leading farmers to rely on fungicide applications. 
These interventions are predominantly reactive and applied only after visible signs of infection are observed \cite{avelino2015impact}. 
This delayed response often allows the pathogen to spread uncontrollably, exacerbating yield losses and increasing production costs. 
Moreover, while fungicides can offer short-term relief, their widespread use poses significant challenges. 
Repeated applications contribute to the development of fungicide-resistant pathogen strains, reducing long-term efficacy and necessitating the use of higher doses or new chemical formulations \cite{brent2003fungicide, jeger2008adaptation}. 
Additionally, fungicides can negatively impact non-target organisms, disrupt ecological balances and lead to environmental contamination, particularly in regions where excessive use results in soil and water pollution \cite{vanbruggen2016impact}. 
To reduce fungicide dependency and enhance early detection of coffee leaf rust, sustainable disease management strategies are crucial. 
An integrated approach, combining predictive modelling, resistant cultivars, optimized cultural practices and biocontrol agents offers a more effective solution. 

Biological control is a sustainable and environmentally friendly alternative to chemical fungicides. 
It involves the use of natural antagonists to suppress  reproductive structures of the pathogens and reduce the progression of the disease.
Among the most studied biocontrol agents, the fungus \textit{Lecanicillium lecanii} has shown significant potential. 
This mycoparasitic fungus parasitizes uredospores of \textit{Hemileia vastatrix}, thereby reducing their germination and viability \cite{agronomy11091865,vandermeer2009evidence}. 
Another promising agent is \textit{Calonectria hemileiae}, which has been shown to decrease the severity of coffee leaf rust by over 90\% through a combination of spore inhibition and enhanced plant resistance \cite{salcedo2021elucidating}. 
Several insects such as \textit{Mycodiplosis} can feed on CLR spores \cite{henk2011mycodiplosis}. 
Other investigations on biocontrol focused on antagonist bacteria such as \textit{Bacillus} species make plants more resistant to fungal infections \cite{daivasikamani2009biocontrol,heydari2010biocontrol, singh1984bacillus}.
In all these biocontrol studies, the dose and time of application of the bacteria, hyperparasites or predators affected their effectiveness in controlling coffee leaf rust development.

Mathematical models have been instrumental in advancing our understanding of plant disease dynamics and evaluating control strategies. 
Deterministic models, such as those based on compartmental frameworks (e.g., susceptible-infected-removed models), have been widely used to study crop-pathogen interactions \cite{madden2007study, mailleret2009semi, nundloll2010impulsive, Djuikem2025busseola}. 
These models effectively capture average disease trends and have been extended to analyse the spread of fungal diseases, including CLR, under various environmental and management conditions \cite{djuikem2021modelling,meng2010plantdisease}. 
In particular, for coffee leaf rust (CLR), one of the authors developed a biocontrol model based on existing literature that promotes biological control strategies \cite{djuikem2023impulsive}. 
Their model demonstrated that the effectiveness of a predator-based biocontrol strategy depends on key factors such as consumption and mortality rates of the predator\cite{setiawati2021variability}.
A semi-numerical analysis of this impulsive control model revealed that biocontrol could significantly reduce CLR under appropriate conditions. 
Furthermore, their study examined how the success of biocontrol implementation is influenced by predator characteristics, particularly mortality rates \cite{HAJIANFOROOSHANI2023105099}, within a framework where the total annual release of predators was fixed, varying only the frequency of release. 
This approach allowed for a comparison between different release strategies under a constant yearly budget allocated to biocontrol agents. 
The study evaluated scenarios with both high and low predator mortality rates, representing different environmental conditions and predator characteristics. 
High mortality was linked to specialist predators, such as \textit{Mycodiplosis}, which exclusively consume CLR uredospores and rapidly perish in the absence of the pathogen \cite{henk2011mycodiplosis}. 
Conversely, low mortality was associated with generalist predators in a resource-abundant environment, such as \emph{Lecanicillium lecanii}, a hyperparasite of CLR that also targets plant-parasitic nematodes \cite{agronomy11091865, vandermeer2009evidence}. Both mortality scenarios are realistic and can be observed in field conditions.

Despite the extensive body of research on biocontrol strategies, existing models often overlook the specific context of smallholder farmers, who constitute the majority of coffee producers worldwide. 
Small-scale farmers primarily depend on cultural practices as their principal disease management strategy \cite{ DUPRE2022105918}. 
The application of  biological control agents is typically a reactive approach, initiated only when visible symptoms of infection or pest infestation manifest in the plantation.
In this study, we extend the existing literature, particularly \cite{djuikem2023impulsive}, by assuming that biocontrol measures are implemented when a defined infection threshold is reached rather than at set times.  
This allows to shift the focus toward the decision-making processes of smallholder farmers. 

This paper is organised as follows. 
Section~\ref{sec:ODE-IDE} is devoted to the formulation of ordinary differential equation (ODE) and impulsive differential equation (IDE) models and their analysis. 
It also presents the control strategies and different scenarios with numerical simulations.  
In Section~\ref{sec:CTMC-MBPA}, we introduce the continuous-time Markov chain (CTMC) associated to the ODE to focus on the early propagation of CLR in the plantation. 
We analyse the CTMC using a multitype branching process approximation (MBPA) and consider scenarios of disease extinction even without biocontrol. 
Section~\ref{sec:Discussion} concludes the paper with a discussion of the main results and perspectives for future work.

\section{ODE and IDE models and their analysis}
\label{sec:ODE-IDE}
We start by investigating the ordinary differential equations (ODE) and impulsive differential equations (IDE) models.
In Section~\ref{sec:ODE-model}, we consider the base ODE model for coffee leaf rust and a biocontrol agent.
Then, in Section~\ref{sec:IDE-yearly}, we add periodic (yearly) cultivation practices. 
Finally, in Section~\ref{sec:IDE-control}, we add threshold-based release of biocontrol agents to the model.
In the three cases, we first formulate the model, then conduct mathematical or computational analyses.

\subsection{The base ODE model for coffee rust and predators}
\label{sec:ODE-model}
The model describes the dynamics of CLR in a coffee plantation, focusing on the interactions between four compartments: susceptible leaves $S$, infected leaves $I$, fungal spores $U$ and predators used as biocontrol agents $P$. 

The population $S$ of susceptible leaves increases at a constant recruitment rate \(\Lambda\), representing natural leaf growth. 
Spores land on leaves at the rate $\omega\nu SU$; when conditions are favourable, this leads to germination of spores at the same rate.
This process is destructive of spores, leading to mortality of spores at the rate $\nu SU$.
Fungal spores are produced by infected leaves at a \emph{per capita} rate \(\gamma\). 
All compartments are subject to natural mortality at the \emph{per capita} rates \(\mu\) for leaves and \(\mu_U\) for spores. 
Infected leaves are subject to an additional disease-induced mortality at the \emph{per capita} rate \(d\). 
Natural predation by predators, represented by \(\delta UP\), where \(\delta\) is a consumption rate of spores by predators. 
The predator population grows due to the consumption of fungal spores at the rate \(\eta \delta UP\), where \(\eta\) is the conversion efficiency of spores into predator biomass. 
However, predators also experience natural mortality at a rate \(\mu_P P\).
The model then takes the following form,
\begin{subequations}
\label{sys:ode-model}
\begin{align}
    \dot S &= \Lambda- \omega \nu SU-\mu S \label{sys:ode-model-dS} \\
    \dot I & = \omega \nu S U  -(\mu+d) I \label{sys:ode-model-dI}  \\
    \dot U &=  \gamma  I - \nu SU -\mu_U U -\delta UP \label{sys:ode-model-dU} \\
    \dot P & = \eta \delta UP-\mu_P P. \label{sys:ode-model-dP} 
\end{align}
\end{subequations}
System \eqref{sys:ode-model} is considered with nonnegative initial conditions.
Parameters used in \eqref{sys:ode-model} are summarised in Table~\ref{tab:model-parameters}.

\begin{table}[H]
\centering
\caption{Parameters of the model.}
\label{tab:model-parameters}
\begin{tabular}{cll}
\toprule
\textbf{Parameter} & \textbf{Description} & \textbf{Value} \\ \midrule
$\Lambda$ & Recruitment rate of susceptible leaves & 3 leaves/day \\
$\omega$ & Germination rate & 0.05 leaf/spores \cite{Rayner1961}\\ 
$\nu$ & Infection rate coefficient & 0.009/leaf/day \\ 
$\mu$ & Natural death rate of $S$ and $I$ & 0.014/day\\ 
$d$ & Disease-induced death rate of $I$ & 0.056/day \\ 
$\gamma$ & Production rate of spores by $I$ & 5 spores/leaf/day  \cite{Bock1962} \\ 
$\mu_U$ & Natural death rate of spores & 0.035/day \\ 
$\delta$ & Consumption rate of spores by predators & 0.3 spores/predator/day \\ 
$\eta$ & Biomass conversion of spores to predators & 0.005 predators/spore \\ 
$\mu_P$ & Natural death rate of predators &  0.01/day\\ 
$T$ & Length of the year & 365 days \\
\bottomrule
\end{tabular}
\end{table}

\subsubsection{Basic properties of the ODE model}
\label{subsec:basic-properties-ODE}
Solutions of \eqref{sys:ode-model} remain non-negative for all times. 
Indeed, consider any variable \(x\in\{S, I, U, P\}\). 
Suppose by contradiction that \(x\) becomes negative at some time \(t^*>0\). 
A quick analysis shows that \(\dot x(t)\geq 0\), so $x(t^*)$ cannot become negative.

\subsubsection{Equilibria and their stabilities}
\label{subsec:EP-ODE}
The set $\{P=0\}$ is positively invariant under the flow of \eqref{sys:ode-model}.
On this set, \eqref{sys:ode-model} admits two equilibria: the disease-free equilibrium (DFE) $\bE^0= (S^0,0,0,0)$, where $S^0=\Lambda/\mu$, and a predator-free endemic equilibrium (PFEE) $\bE^\star$ whose expression is made explicit later.

Note that although we are at this point working on the invariant set $\{P=0\}$, results concerning the local stability or instability of the DFE apply to the entire system \eqref{sys:ode-model} as computations on $\{P=0\}$ are exactly the same as those on the whole $\IR_+^4$, since the next generation matrix method of \cite{VdDWatmough2002} only focuses on infected variables $(I,U)$.
Results concerning the predator-free endemic equilibrium differ: while the PFEE itself is the same on $\IR_+^4$ and $\IR_+^4\setminus\{P=0\}$, determining the local asymptotic stability in $\IR_+^4\setminus\{P=0\}$ requires to use one less variable ($P$) as does the local asymptotic stability in  $\IR_+^4$.

Bearing this in mind, we now proceed to compute the basic reproduction number $\R_0$, whose influence, as we just noted, extends to the whole of \eqref{sys:ode-model}.
To determine the matrices used in the computation of the basic reproduction number using the next generation matrix method of \cite{VdDWatmough2002}, order infected and spores variables as $(I,U)$. 
The basic reproduction number is $\mathcal{R}_0=\rho(F V^{-1})$, where $\rho$ is the spectral radius and
\[F=\begin{pmatrix}
	\omega \nu S^0 & 0\\
	0 &0
\end{pmatrix} \text{ and }  V= \begin{pmatrix}
	\mu+d & 0\\
	\gamma & \nu U S^0 -\mu_U 
\end{pmatrix},
\]
i.e., 
\begin{equation}
	\label{eq:R0}
	\mathcal{R}_0=\frac{\gamma \nu \omega S^0}{(d + \mu)( S^0 \nu + \mu_U)}.
\end{equation}
Hypotheses of \cite[Theorem 2]{VdDWatmough2002} are easily checked and we deduce that $\R_0=1$ acts as a threshold for the local asymptotic stability of the disease-free equilibrium.
Using the proof in Appendix~\ref{sec:appen-proof-theo-gas}, the result can in fact be made global, as stated in the following result.

\begin{lemma}\label{lm:LAS-DFE}
    The DFE $\bE^0$ of \eqref{sys:ode-model} is globally asymptotically stable when $\mathcal{R}_0<1$ and unstable otherwise.  
\end{lemma}

We now turn our attention to endemic equilibria of \eqref{sys:ode-model}.
We have the following.
\begin{lemma}
	System \eqref{sys:ode-model} has two endemic equilibria.
	The first is the predator-free endemic equilibrium (PFEE)
	\begin{equation}
		\label{eq:PFEE}
		\bE^\star=(S^\star,I^\star,U^\star,P^\star)
		=\left(
		\frac{\Lambda}{\mu+\omega\nu U^\star},
		\frac{\Lambda\omega\nu U^\star}{(\mu+\omega\nu U^\star)(\mu+d)},
		\frac{(\Lambda\nu+\mu\mu_U)(\mathcal{R}_0-1)}{\mu_U \omega \nu },
		0
		\right),
	\end{equation}
	which is an equilibrium in $\IR_+^4\setminus\{P=0\}$ that is biologically relevant when $\R_0>1$.
	The second is the endemic equilibrium point (EEP) $\bE_\star=(S_\star,I_\star,U_\star,P_\star)$, where
	\begin{equation}
		\label{eq:EEP}
		\bE_\star =
		\left(
		\frac{\Lambda}{\mu+\omega \nu U_\star},
		\frac{\Lambda\omega\nu U_\star}{(\mu+\omega\nu U_\star)(\mu+d)},
		\frac{\mu_P}{\eta \delta },
		\frac{\eta \delta(\Lambda \nu +\mu \mu_U)(\mathcal{R}_0-1)-\mu_P \mu_U\omega\nu}{\delta(\eta \delta \mu+\mu_P\omega\nu)}
		\right),
	\end{equation}
	which is biologically relevant if and only if
	\begin{equation}
		\label{eq:EEP-relevant}
		\eta\delta(\Lambda\nu+\mu\mu_U)(\R_0-1)-\mu_P\mu_U\omega\nu>0.
	\end{equation}
	In particular, when $\R_0<1$, the EEP is not biologically relevant.
\end{lemma}

\begin{proof}
The expression for the PFEE is shared between the system on $\IR_+^4\setminus\{P=0\}$ and that on $\IR_+^4$.
When $P^\star=0$, we must solve
\begin{subequations}
	\begin{align}
		0 &= \Lambda- \omega \nu SU-\mu S,  \label{PFE:S} \\
		0 & = \omega \nu S U  -(\mu+d) I, \label{PFE:I} \\
		0 &=  \gamma  I - \nu SU -\mu_U U. \label{PFE:U} 
	\end{align}
\end{subequations}
From \eqref{PFE:S}, $S=\Lambda/(\mu+\omega\nu U)$. Substituting into \eqref{PFE:I} gives
\[
I=\frac{\Lambda\omega\nu U}{(\omega\nu U+\mu)(\mu+d)}.
\]
Substituting the expressions for $S$ and $I$ into \eqref{PFE:U} gives
\[
\frac{\Lambda\gamma\omega\nu U}{(\mu+\omega\nu U)(\mu+d)}
-\frac{\Lambda\nu U}{\mu+\omega\nu U}-\mu_UU = 0.
\]
Since $\mu+\omega\nu U$ cannot be zero, $U^\star$ are the zeros of
\[
\Lambda\gamma\omega\nu U-(\mu+d)\Lambda\nu U-\mu_U(\mu+\omega\nu U)(\mu+d)U=0.
\]
Rewrite this polynomial as
\[
\mu_U(\mu+d)\omega\nu U^2+(\mu_U\mu(\mu+d)+(\mu+d)\Lambda\nu-\Lambda\gamma\omega\nu)U = 0.
\]
One of the roots is $U^\star=0$, giving the DFE, while the other is
\[
U^\star = \frac{\Lambda\gamma\omega\nu-\mu_U\mu(\mu+d)-(\mu+d)\Lambda\nu}{\mu_U(\mu+d)\omega\nu}=\frac{(\mathcal{R}_0-1)(\mu_U \mu+\Lambda \nu)}{\mu_U \omega \nu }.
\]
Therefore, the predator-free endemic equilibrium $\bE^\star$ takes the form \eqref{eq:PFEE}.
Note that the PFEE only makes sense biologically when $\R_0>1$, in which case all its components except $P^\star$ are positive.

Assume now that $P>0$, which, to distinguish from the other two cases, we denote $P_\star$.
Let $\bE_\star=(S_\star,I_\star,U_\star,P_\star)$ be the endemic equilibrium of system~\eqref{sys:ode-model} obtained when assuming that $P_\star>0$.
At this equilibrium, $U_\star=\mu_P/(\eta\delta)>0$ and we obtain the remaining components as functions of $U_\star$ as given by \eqref{eq:EEP}.
Finally, condition \eqref{eq:EEP-relevant} is easily obtained by requiring that $P_\star>0$.
\end{proof}

We use the notation $\bE_\star$ to indicate that, at that equilibrium point, there is ``less disease'' than at $\bE^\star$.
As noted earlier, establishing the local asymptotic stability of $\bE^\star$ differs whether we are working on the full model \eqref{sys:ode-model} or on \eqref{sys:ode-model} restricted to the set $\IR_+^4\setminus\{P=0\}$.
Since we are interested in solutions for the full system, we consider the complete Jacobian matrix of \eqref{sys:ode-model}, which takes the form
\begin{equation}
	\label{eq:jacobian}
	J(S,I,U,P)
	=\begin{pmatrix}
		-\omega\nu U-\mu & 0 & -\omega\nu S & 0 \\
		\omega\nu U & -(\mu+d) & \omega\nu S & 0 \\
		-\nu U & \gamma & -\nu S-\delta P -\mu_U & -\delta U \\
		0 & 0 & \eta\delta P & \eta\delta U-\mu_P
	\end{pmatrix}.
\end{equation}
This matrix is difficult to study and we mostly consider the stability of $\bE^\star$ and $\bE_\star$ numerically; see below. However, an important observation is that, at $\bE^\star$, \eqref{eq:jacobian} takes the form
\[
J(E^\star)=\begin{pmatrix}
	-\omega\nu U^\star-\mu & 0 & -\omega\nu S^\star & 0 \\
	\omega\nu U^\star & -(\mu+d) & \omega\nu S^\star & 0 \\
	-\nu U^\star & \gamma & -\nu S^\star-\mu_U & -\delta U^\star \\
	0 & 0 & 0 & \eta\delta U^*-\mu_P
\end{pmatrix},
\]
which, as a block upper triangular matrix, has the obvious eigenvalue $\lambda=\eta\delta U^*-\mu_P$.
Substituting $U^\star$ as given by \eqref{eq:PFEE},
\[
\lambda = 
\eta\delta\frac{(\Lambda\nu+\mu\mu_U)(\mathcal{R}_0-1)}{\mu_U \omega \nu }-\mu_P
= \frac{\eta\delta(\Lambda\nu+\mu\mu_U)(\R_0-1)-\mu_P\mu_U\omega\nu}{\mu_U\omega\nu}.
\]
This expression is positive when the condition \eqref{eq:EEP-relevant} for biological relevance of $\bE_\star$ is satisfied and nonpositive when it is not.
As a consequence, whenever $\bE_\star$ is biologically relevant, $\bE^\star$ is unstable.

\begin{figure}[htbp]
    \centering
        \centering
        \includegraphics[width=0.8\textwidth]{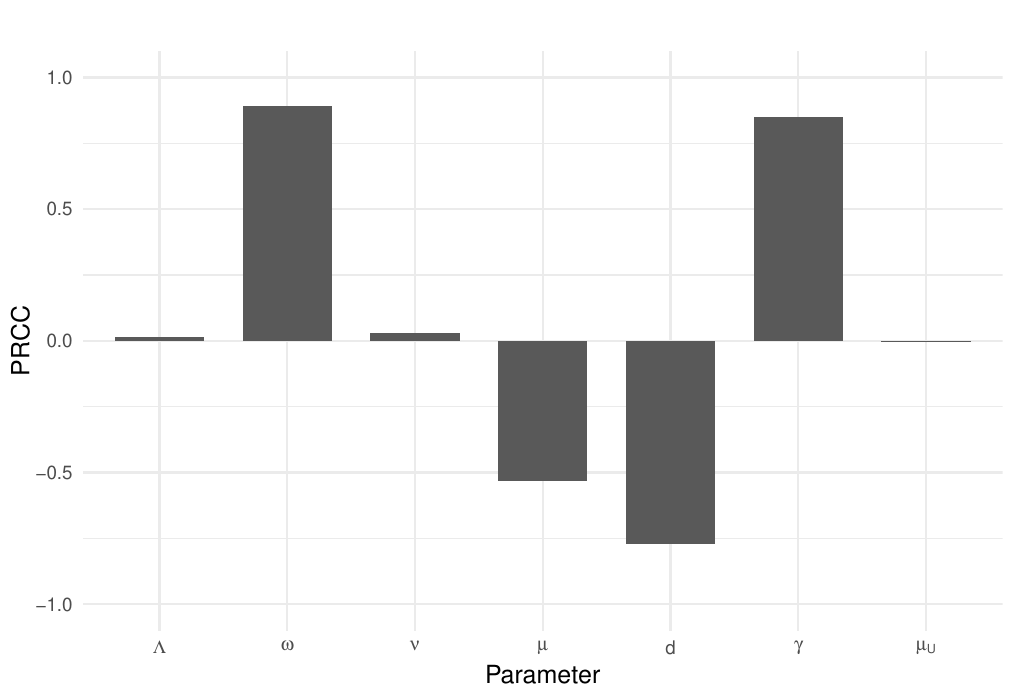} 
    \caption{Partial Rank Correlation Coefficient (PRCC) sensitivity analysis to model parameters of the basic reproduction number \(\mathcal{R}_0\).}
    \label{fig:prcc-analysis}
\end{figure}

We suspect that when $\R_0>1$ and $\bE_\star$ is not biologically relevant, $\bE^\star$ is locally asymptotically stable.
However, the expression for the remaining eigenvalues of $J(\bE^\star)$ is too complicated to make any sense of and we were unable to mobilise spectral theoretic results to provide an answer.
We did investigate the situation numerically. 
In all the computations involved in the sensitivity analyses in Figure~\ref{fig:prcc-analysis}, we computed the values of the equilibria and evaluated their stability numerically.
Among the hundreds of thousand of points in parameter space that we sampled, we always observed that when $\bE^\star$ is present and $\bE_\star$ absent, all eigenvalues of $J(\bE^\star)$ had negative real parts.

Let us now consider the effect of parameters on $\R_0$. 
The partial rank correlation coefficients (PRCC) results presented in Figure~\ref{fig:prcc-analysis} show the most influential parameters affecting  the basic reproduction number \( \mathcal{R}_0 \).
From Figure~\ref{fig:prcc-analysis}, the spore production rate \( \gamma \) and the germination rate \(\omega\) exhibit the highest positive correlations with \( \mathcal{R}_0 \), while the mortality rate due to CLR \( d \) and the natural death rate \( \mu \) show strong negative correlations with \( \mathcal{R}_0 \).

\subsection{The IDE model for a yearly cultivation cycle}
\label{sec:IDE-yearly}

Before implementing control measures, we refine model~\eqref{sys:ode-model} based on the work of \cite{djuikem2023impulsive} to consider a plantation scenario with a single annual harvest as is typical in Cameroon, the setting inspiring the work. At the time of harvest, in addition to crop collection, complementary cultural practices such as pruning to remove old leaves, weed management, and fertilization are implemented to improve plant health and reduce the number of fungal spores in the plantation.
The result is a model with an impulse at the harvest period, which we consider to be a single event each year. 
The impulses occur at times $nT$ and have impact $\varphi_k$, $k \in\{S, I, U, P\}$, where $\varphi_k \in (0,1)$. 
The system is described by the impulsive model~\eqref{sys:imp} below.

\begin{equation}\label{sys:imp}
\begin{array}{rl}
    \text{When } t \in(nT,(n+1)T], &  
    \left\{
    \begin{aligned}
        \dot S &= \Lambda - \omega \nu U S - \mu S, \\
        \dot I &= \omega \nu S U - (\mu + d) I, \\
        \dot U &= \gamma I - \nu U S - \mu_U U - \delta UP, \\
        \dot P &= \eta \delta UP - \mu_P P.
    \end{aligned}
    \right. \\[1.5cm]
    \text{When } t = nT, & 
     \left\{
    \begin{aligned}
        S(nT^+) &= \varphi_S S(nT), \\
        I(nT^+) &= \varphi_I I(nT), \\
        U(nT^+) &= \varphi_U U(nT), \\
        P(nT^+) &= \varphi_P P(nT).
    \end{aligned}
    \right.
\end{array}
\end{equation}

\subsubsection{Basic properties of the IDE model}\label{subsec:basic_proper}
We saw in Section~\ref{subsec:basic-properties-ODE} that solutions to \eqref{sys:ode-model} are nonnegative and bounded for all times.
We just need to check here that adding impulses does not change this.

The impulsive conditions at \(t = nT\) are given by non-negative multiplicative factors \(\varphi_k \in (0,1)\). 
If at time \(nT\) the variables are non-negative, the immediate post-impulse values remain non-negative. 
This implies that with non-negative initial conditions, state variables remain non-negative at all times.
Let us now show that solutions are uniformly bounded.
Consider the the total number of leaves \(N = S+I\). 
Adding the first two differential equations of~\eqref{sys:imp}, we obtain
   \[
   \dot N = \Lambda - \mu N - dI \leq  \Lambda - \mu N .
   \]
With non-negative initial condition $N(0)$ and without impulses, \(N(t)\) never exceeds \(\max(\Lambda/\mu,N(0))\).
Now, accounting for impulses,
\[
N(nT^+) = \varphi_S S(nT) + \varphi_I I(nT) \leq S(nT) + I(nT) = N(nT),
\]
since \(0 \leq \varphi_S,\varphi_I \leq 1\). 
Thus, impulses reduce the total number of leaves \(N\). 
Iterating this argument over each interval \([nT,(n+1)T]\), we find that \(N(t)\) remains bounded above by a constant \(\Gamma_N := \max\left(\Lambda/\mu,N(0)\right)\).
Since \(S,I \geq 0\) and \(S+I=N\), both \(S(t)\) and \(I(t)\) are individually bounded above by \(\Gamma_N\). 
Suppose that $N(0)<\Lambda/\mu$; substitute this upper bound into the equation of spores, giving
\begin{equation*}
\left\{
\begin{aligned}
&\dot{U} \leq \gamma \frac{\Lambda}{\mu} - \nu U S - \mu_U U \leq \gamma \frac{\Lambda}{\mu} - \mu_U U, \\[6pt]
&U(nT^+) = \varphi_U U(nT).
\end{aligned}
\right.
\end{equation*}
Solving the above equation, we obtain $U(t)\leq \Gamma_U:= \max(\Lambda \gamma/(\mu\mu_U),U(0))$.

Concerning the population of biocontrol agents, remark that there holds that $\dot P \leq (\eta \delta \Gamma_U- \mu_P) P$.
This implies that
\[
P(t) \leq P(t_0)e^{\left(\eta \delta \Gamma_U- \mu_P\right)(t-t_0)},
\quad t\geq t_0.
\]
Taking the impulses at $t=nT$ into account, this means that
\[
P(t) \leq P(nT^+)e^{\left(\eta \delta \Gamma_U- \mu_P\right)(t-nT)},
\quad t\in[nT,(n+1)T].
\]
As a consequence, if $\eta \delta \Gamma_U- \mu_P<0$, i.e., $\Gamma_U < \mu_P/\eta \delta$, then since $P(nT^+) = \varphi_P P(nT)$, there holds that $P(t)\leq P(0)\varphi_P^{n+1}$ and the population of biocontrol agents remains bounded. 
The population also remains bounded if $P(0)=0$, as is used in most of our numerical applications.

Finally, smoothness of the right side of \eqref{sys:imp} guarantee the existence and uniqueness of the solutions of this system.

\subsubsection{Periodic solution and stability}

System~\eqref{sys:imp} can be analyzed using the approach outlined in \cite[Section 2]{djuikem2023impulsive}, which we follow. See that paper for details.
In this framework, the periodic disease-free solution (PDFS) is given by
\[
\bE^T(t) = (S^T(t), 0, 0, 0),
\]
where
\begin{equation}
    S^T(t) = \frac{\Lambda}{\mu} \left[ 1 - \frac{(1 - \varphi_S) e^{\mu T}}{e^{\mu T} - \varphi_S} e^{-\mu(t - nT)} \right].
\end{equation}
To evaluate the stability of the PDFS, we calculate the spectral radius of the monodromy matrix associated with \eqref{sys:imp} using Floquet theory. 
This yields
\begin{equation}\label{eq:R}
    \mathcal{R} = \frac{\varphi_I}{2\beta} 
\left[ (\beta + k_1 - k_2)e^{\lambda_1 T} + (\beta - k_1 + k_2)e^{\lambda_2 T} \right],
\end{equation}
where parameters are defined as
\begin{align*}
k_1 &= \mu + d, \\
k_2 &= \nu S^0 + \mu_U, \\
\alpha &= k_1 + k_2, \\
\beta &= \sqrt{(k_1 - k_2)^2 + 4\gamma \omega \nu S^0}, \\
\lambda_1 &= -\frac{\alpha}{2} - \frac{\beta}{2}, \\
\lambda_2 &= -\frac{\alpha}{2} + \frac{\beta}{2}.
\end{align*}
The stability of the PDFS is characterized by the following result.
\begin{lemma}\label{lm:gloal-sta-PDFS}
    The PDFS \( \bE^T(t) \) is locally asymptotically stable if \( \mathcal{R} < 1 \) and unstable if \( \mathcal{R} > 1 \).
\end{lemma}
The proof of Lemma~\ref{lm:gloal-sta-PDFS} follows exactly that of \cite[Lemma 3]{djuikem2023impulsive} and is therefore omitted.
The following result, whose proof is found in Appendix~\ref{sec:appen-proof-theo-gas}, links the global stability of the PDFS $\bE^T(t)$ of \eqref{sys:imp} to the local asymptotic stability of the DFE $\bE^0$ of the ODE \eqref{sys:ode-model}.
\begin{theorem}\label{thm:gloal-sta-PDFS}
    The PDFS \( \bE^T(t) \) is globally asymptotically stable (GAS) if \( \mathcal{R}_0 < 1 \) and unstable otherwise.
\end{theorem}

We can easily observe that \(\R_0 < 1\implies\R < 1\). Indeed, since \(\lambda_1 < 0\) and \(\R_0 < 1\) leads to \(\gamma \omega \nu S^0 < k_1 k_2\), it follows that \(\lambda_2 < 0\). Consequently, \(e^{\lambda_1 T} < 0\) and \(e^{\lambda_2 T} < 0\). Substituting these into equation~\eqref{eq:R}, we obtain \(\R < \varphi_I < 1\). This clearly establishes the relation between $\R_0$ and $\R$, which is further confirmed by the fact that, to be GAS, $\bE^T$ must also be LAS.

\subsection{Adding prevalence-based control to the IDE model}
\label{sec:IDE-control}
\subsubsection{The model}

Coffee cultivation worldwide is predominantly managed by small-scale producers. 
Due to limited resources, many of these producers often delay implementing control measures until the situation in their plantations has significantly deteriorated. 

In this section, we investigate a control strategy that activates only when the number of infected leaves exceeds a certain threshold.
Specifically, control is initiated  when \(I(t) = I_s\), where \(I_s\) is some infection threshold.
We evaluate the effectiveness of such a strategy with varying thresholds \(I_s\).

We start with the impulsive model~\eqref{sys:imp}, but add control measures at the time \( t_s \) when \( I(t) = I_s \). 
We assume that predators are released at most once a year.
Let \(\Lambda_P\) denote the quantity of predators released at time \( t_s \).  
Combining this information, we derive a model that incorporates two types of impulses:
\begin{itemize}
    \item periodic impulses that occur at the end of the rainy season (\( t = nT \)), representing the impact of harvest and practical agricultural activities;
    \item non-periodic impulses that occur when \( I(t) = I_{s} \), representing farmer-initiated control actions based on observed infection levels.
\end{itemize}

Let $\Delta x= x(t^+)-x(t)$, the resulting state/time-dependent impulsive control model can be written as follows.
\begin{equation}\label{sys:imp-control}
\begin{array}{rl}
    \text{When } t \neq nT, I\ne I_s, &  
    \left\{
    \begin{aligned}
        \dot S &= \Lambda - \omega \nu U S - \mu S, \\
        \dot I &= \omega \nu S U - (\mu + d) I, \\
        \dot U &= \gamma I - \nu U S - \mu_U U - \delta UP, \\
        \dot P &= \eta \delta UP - \mu_P P,
    \end{aligned}
    \right. \\[1.5cm]
    \text{when } I=I_s, & 
     \left\{
    \begin{aligned}
        \Delta S &= 0, \\
        \Delta I &= 0, \\
        \Delta U &= 0, \\
        \Delta P &= \Lambda_P,
    \end{aligned}
    \right. \\[1.5cm]
    \text{when } t = nT, & 
     \left\{
    \begin{aligned}
        S(nT^+) &= \varphi_S S(nT), \\
        I(nT^+) &= \varphi_I I(nT), \\
        U(nT^+) &= \varphi_U U(nT), \\
        P(nT^+) &= \varphi_P P(nT),
    \end{aligned}
    \right.
\end{array}
\end{equation}
with a nonnegative initial condition $\left(S(0^+),I(0^+),U(0^+),P(0^+)\right)$.
Using the same approach as in Section~\ref{subsec:basic_proper}, we conclude that solutions of \eqref{sys:imp-control} exist and are unique.

\subsubsection{Numerical simulation}
The control approach assumes that predators are released at most once a year if the number of infected leaves becomes larger than a threshold \(I_{s}\). 
This threshold can vary from one farmer to another due to differences in how they perceive and define low or high levels of infection. 
This variation is particularly noticeable among small-scale farmers.
Based on their perception and evaluation of the plantation, farmers decide whether the infection level is high enough to justify applying control measures.
The central question is how the threshold \(I_{s}\) impacts the effectiveness of the control strategy.
Consider the initial condition
\begin{equation}\label{eq:int-cond}
    (S(0^+),I(0^+),U(0^+),P(0^+))=(100,0,200,0).
\end{equation}

\paragraph{Impact of the threshold on biocontrol efficiency for $\mathcal{R}_0=1.5$.} 

\begin{figure}[htbp]
    \centering
    \begin{subfigure}{0.45\textwidth}
        \caption{$I_s=10$.}
        \includegraphics[width=\textwidth]{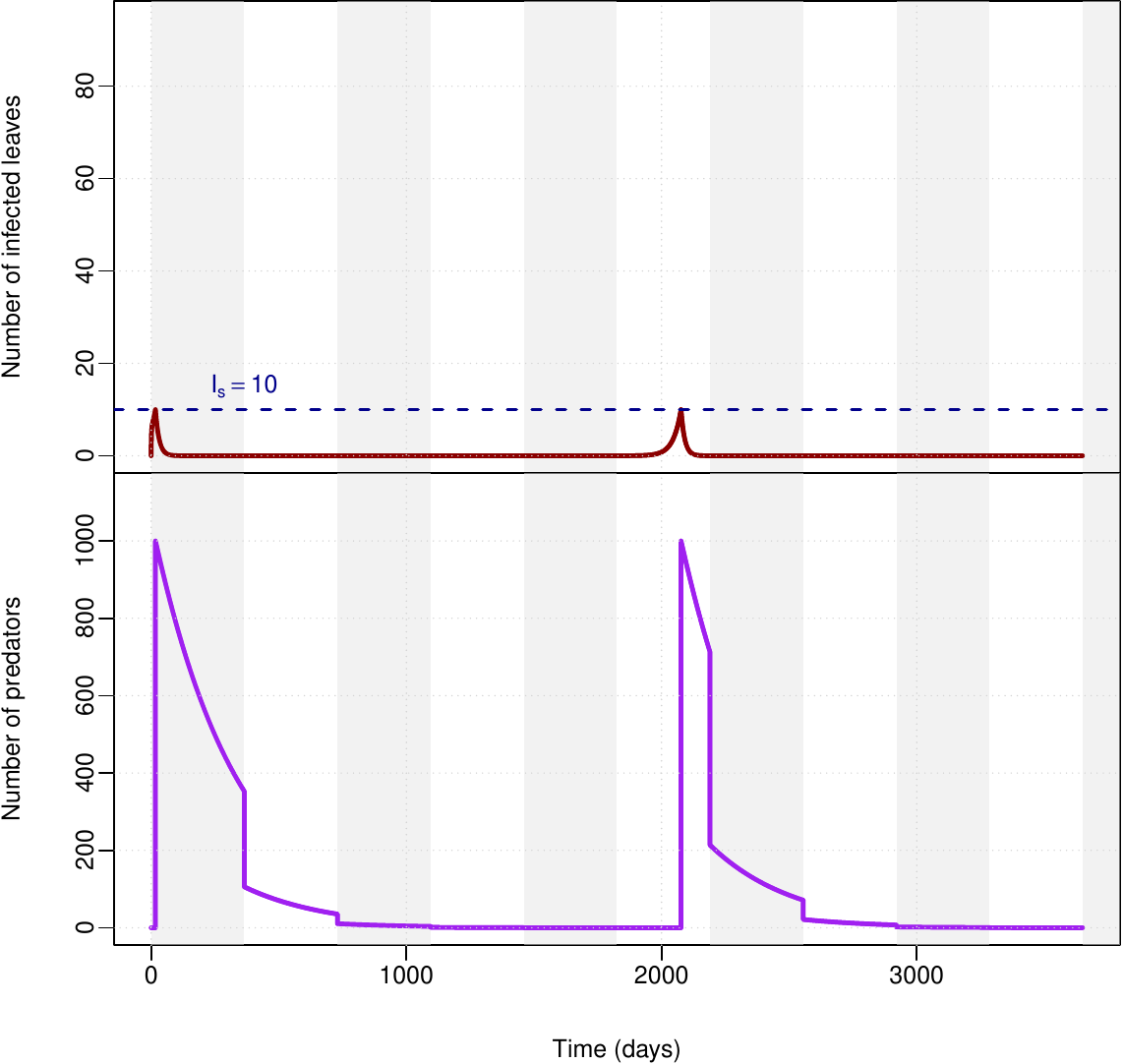}
        \label{fig:R01.5-Is10}
    \end{subfigure}
    \hfill
    \begin{subfigure}{0.45\textwidth}
        \caption{$I_s=50$.}
        \includegraphics[width=\textwidth]{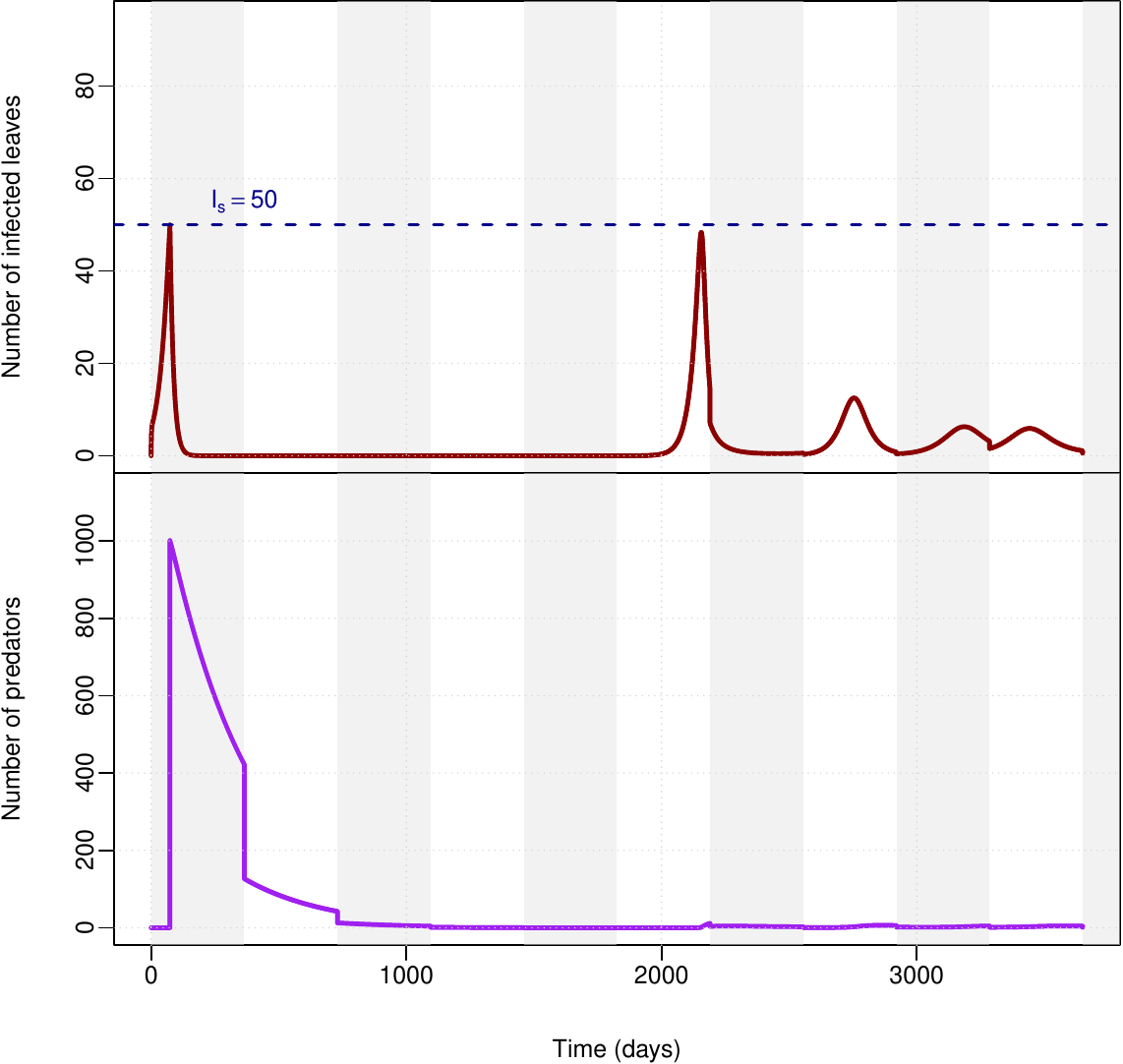}
        \label{fig:R01.5-Is50}
    \end{subfigure} \\
    \begin{subfigure}{0.45\textwidth}
        \caption{$I_s=70$.}
        \includegraphics[width=\textwidth]{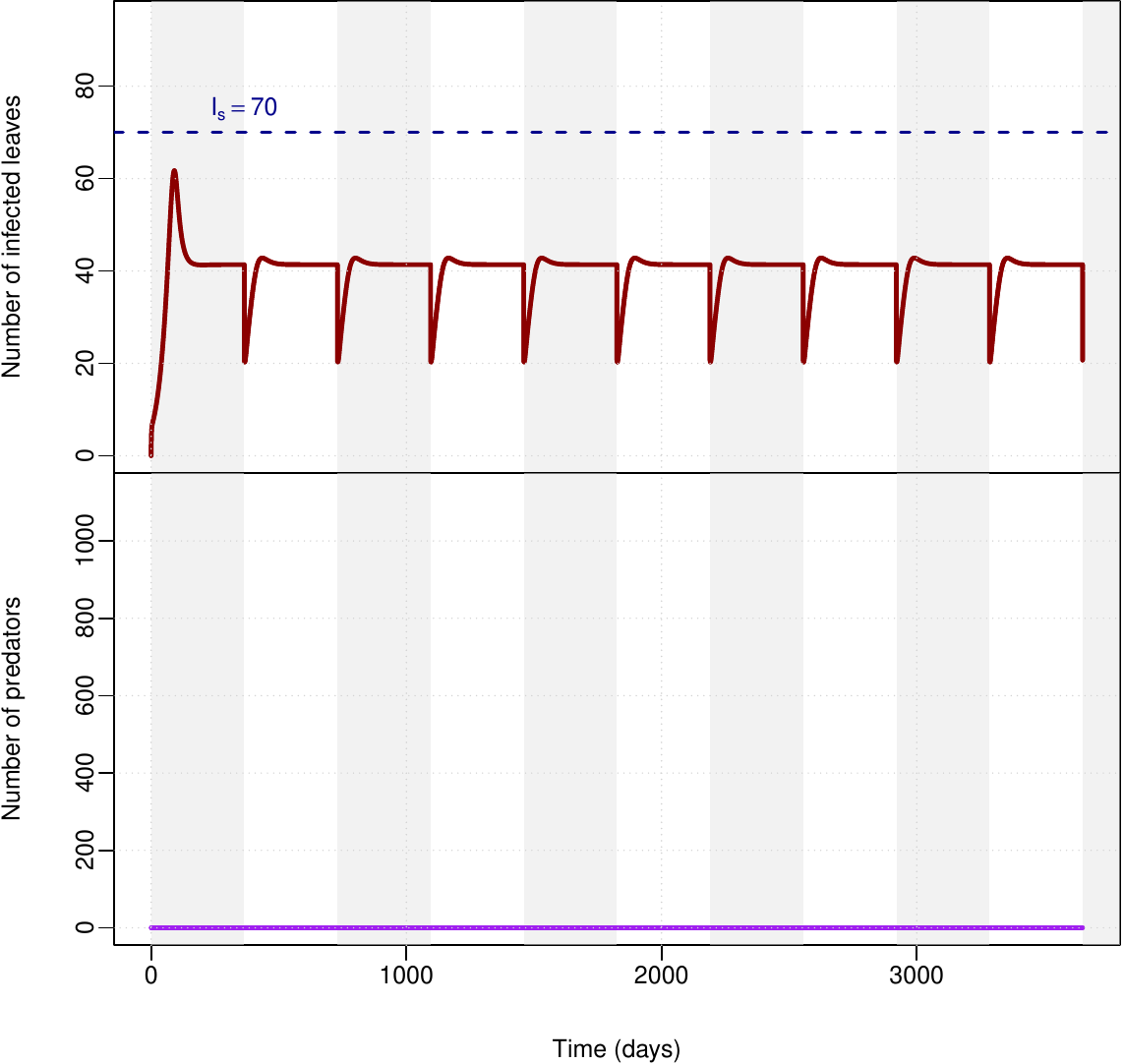}
        \label{fig:R01.5-Is70}
    \end{subfigure}
    \caption{Dynamics of infected leaves $I$ (top panel) and predators $P$ (bottom panel) as a function of the threshold for activation of biocontrol (blue dashed line). 
    Years shown by bands of alternating colours. 
    10-year simulations of model~\eqref{sys:imp} using initial conditions~\eqref{eq:int-cond}. 
    (a) dynamics when $I_s=10$, (b) $I_s=50$ and (c) $I_s=70$. 
    All parameter values as in Table~\ref{tab:model-parameters} except the basic reproduction number $\mathcal{R}_0=1.5$ and $\omega$ computed from $\R_0$.}
    \label{fig:case2-biocontrol}
\end{figure}

Figure~\ref{fig:case2-biocontrol} compares the temporal dynamics of \eqref{sys:imp-control} when the basic reproduction number takes the (small) value \(\mathcal{R}_0=1.5\), with biocontrol applied at different thresholds \(I_s\).

First, when the threshold is low (Figure~\ref{fig:R01.5-Is10}), the number of infected leaves reaches it very quickly, leading to an introduction of predators that keeps the situation under control for quite some time.
However, because infected spores are never completely eradicated, the infection starts over in year 6.
As the threshold increases to $I_s=50$ (Figure~\ref{fig:R01.5-Is50}, the same phenomenon takes place but after resurgence of the infection in year 6, we observe damped oscillations.
These post-release infection peaks are due to \emph{spore-predator dynamics}, when predators initially reduce the number of spores, but as spores decrease, the predator populations decline, allowing infections to return.
Also, because of healthy leaf growth, infection suppression leads to more healthy leaves, which later become susceptible, fuelling new outbreaks.

\begin{figure}[htbp]
    \begin{subfigure}{0.49\textwidth}
        \caption{$\R_0=1.5$.}
        \includegraphics[width=\linewidth]{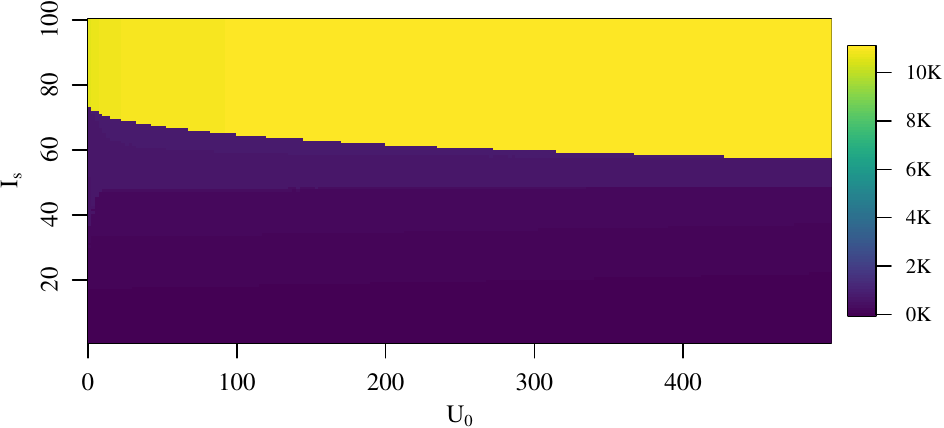}        
        \label{fig:heatmaps-u0-IS-R01.5}
    \end{subfigure}
    \begin{subfigure}{0.49\textwidth}
        \caption{$\R_0=50$.}
        \includegraphics[width=\linewidth]{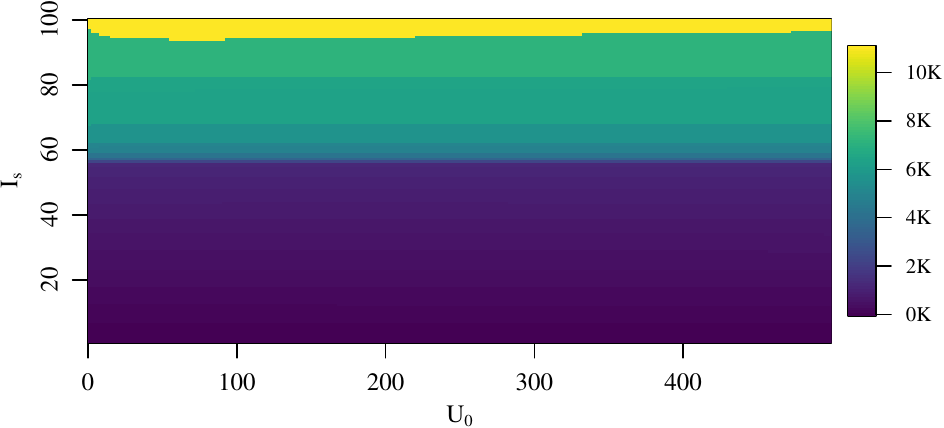}        
        \label{fig:heatmaps-u0-IS-R050}
    \end{subfigure} \\
    \begin{subfigure}{0.49\textwidth}
        \caption{$\mu_P=0.003$.}
        \includegraphics[width=\linewidth]{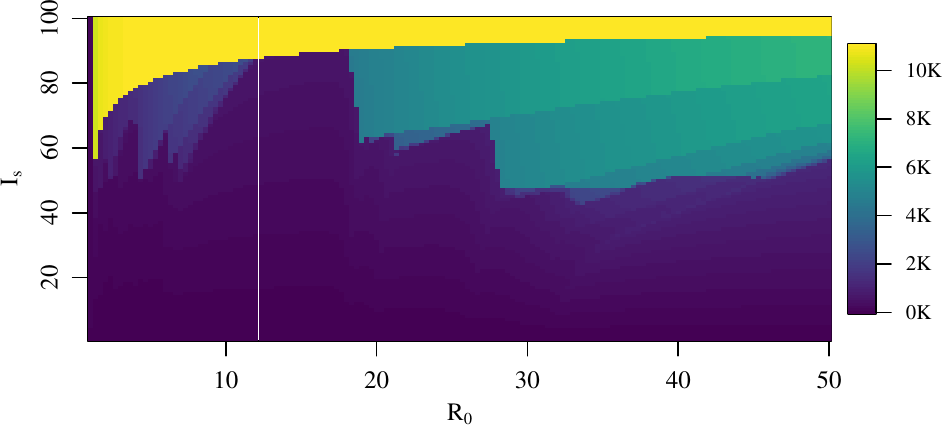}        
        \label{fig:heatmaps-R0-IS-muP0.003}
    \end{subfigure}
    \begin{subfigure}{0.49\textwidth}
        \caption{$\mu_P=0.01$.}
        \includegraphics[width=\linewidth]{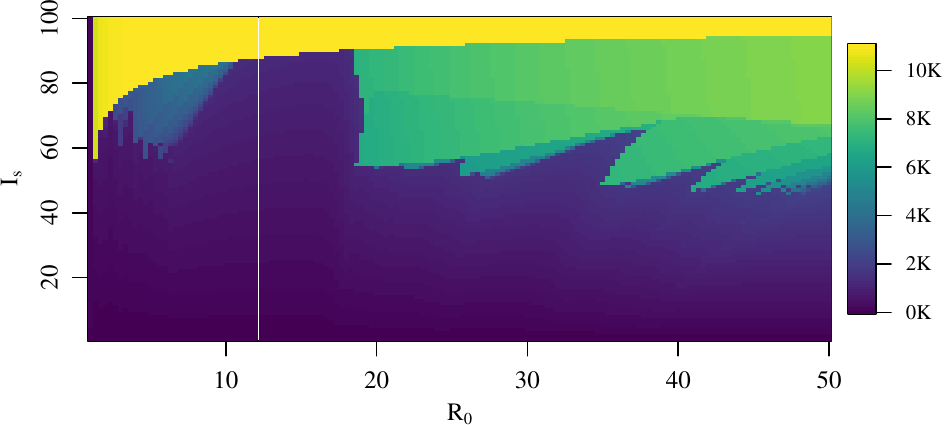}
        \label{fig:heatmaps-R0-IS-muP0.01}
    \end{subfigure} \\
    \begin{subfigure}{0.49\textwidth}
        \caption{$U_0=200$.}
        \includegraphics[width=\linewidth]{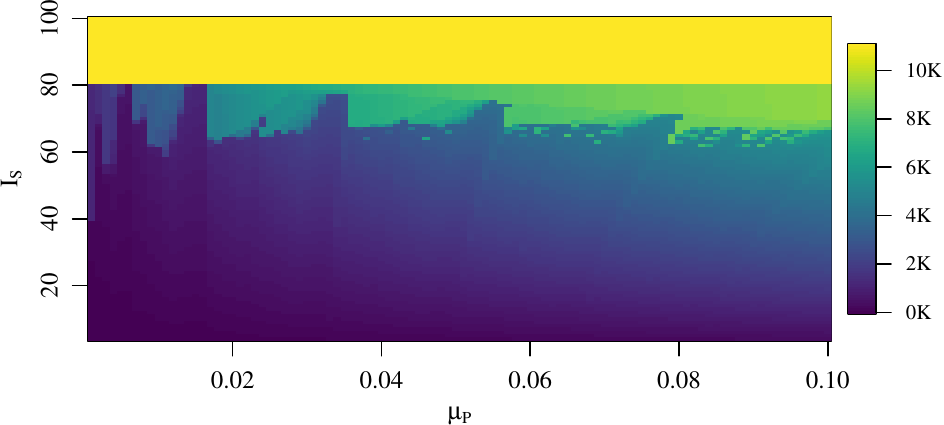}        
        \label{fig:heatmaps-muP-IS-u0200}
    \end{subfigure}
    \begin{subfigure}{0.49\textwidth}
        \caption{Zoom on $U_0=200$.}
        \includegraphics[width=\linewidth]{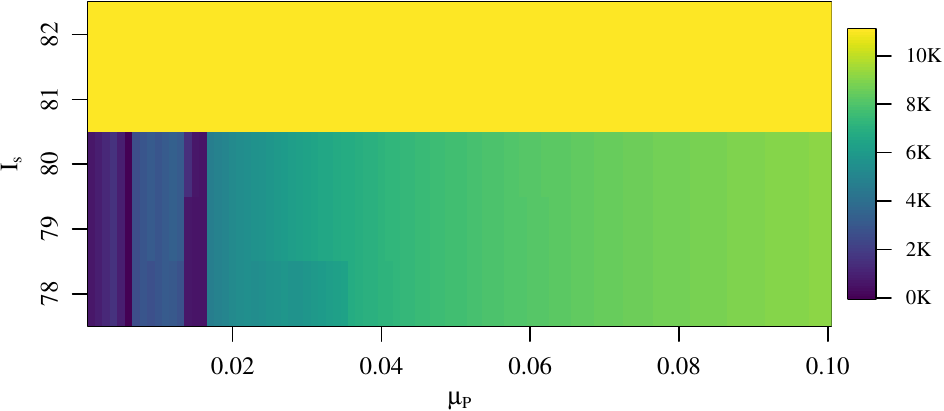}        
        \label{fig:heatmaps-muP-IS-u0200-zoom}
    \end{subfigure}
    \caption{Total number \(I_{\text{total}}(10\text{ years})\) of newly infected leaves as a function of (a,b) the initial number $U_0$ of spores, (c,d) the basic reproduction $\R_0$ and (e,f) the death rate $\mu_P$ of predators, all versus the biocontrol threshold \(I_s\).
    All parameter values are given in Table~\ref{tab:model-parameters}, except for \( \omega \), which is determined for each value of \( \mathcal{R}_0 \).}
    \label{fig:heatmaps}
\end{figure}

In the last figure (Figure~\ref{fig:R01.5-Is70}), we observe the effect of setting too high a threshold: because the threshold is never reached, the solution goes to the periodic endemic equilibrium.
The situation without control ($I_s=\infty$), not shown, is exactly the same as in Figure~\ref{fig:R01.5-Is70}.
See Figure~\ref{fig:heatmaps} later for a measure of the impact of these measures in terms of cumulative infection load.
This emphasizes the need for a lower threshold to ensure timely and effective biocontrol.

\begin{figure}[htbp]
    \centering
    \begin{subfigure}{0.49\textwidth}
        \caption{$\R_0=50$ and $I_s=10$.}
        \includegraphics[width=\textwidth]{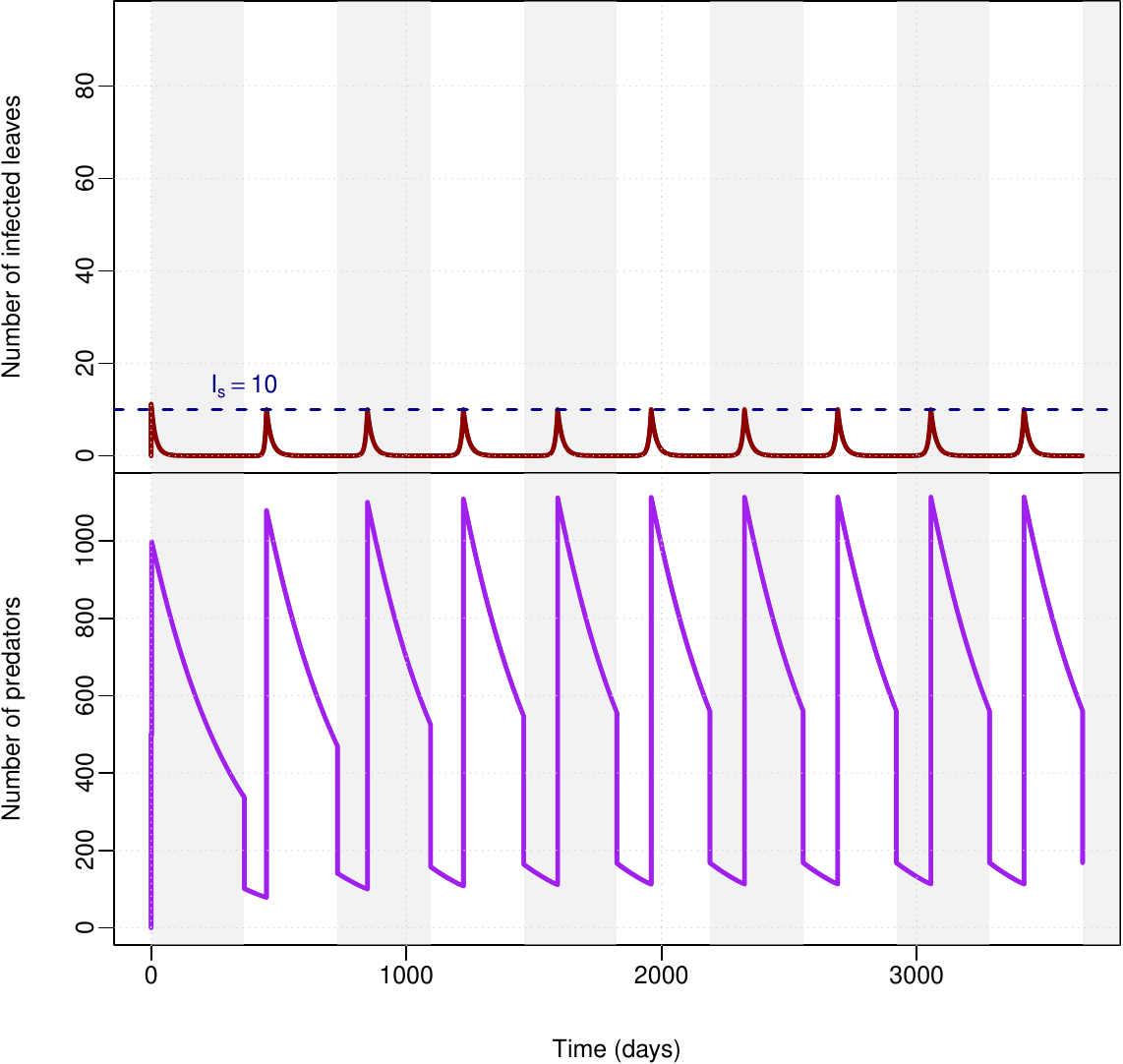}
        \label{fig:R050-Is10}
    \end{subfigure}
    \begin{subfigure}{0.49\textwidth}
        \caption{$\R_0=40$ and $I_s=50$.}
        \includegraphics[width=\textwidth]{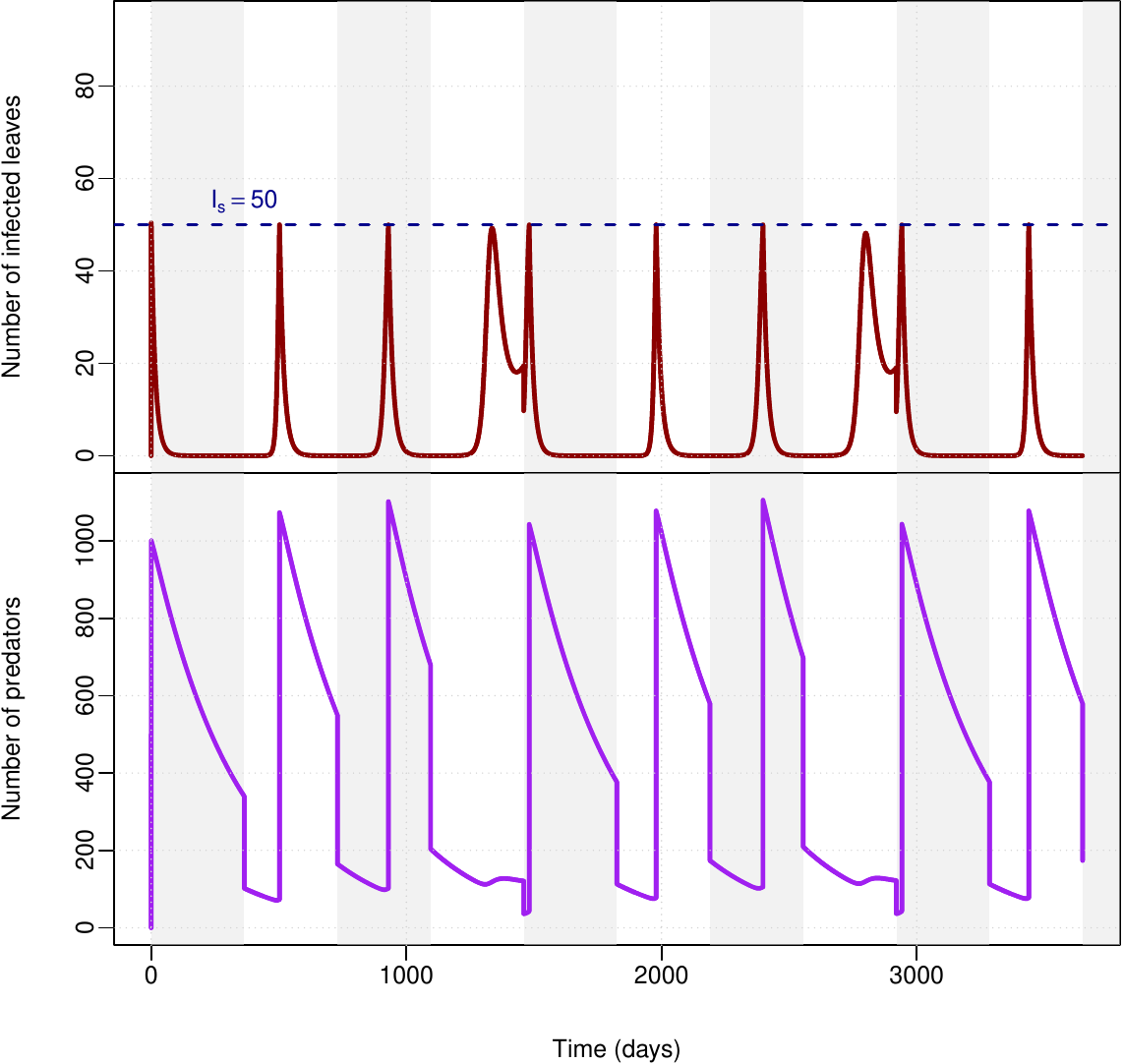}
        \label{fig:R040-Is50}
    \end{subfigure} \\
    \begin{subfigure}{0.47\textwidth}
        \caption{$\R_0=40$ and $I_s=60$.}
        \includegraphics[width=\textwidth]{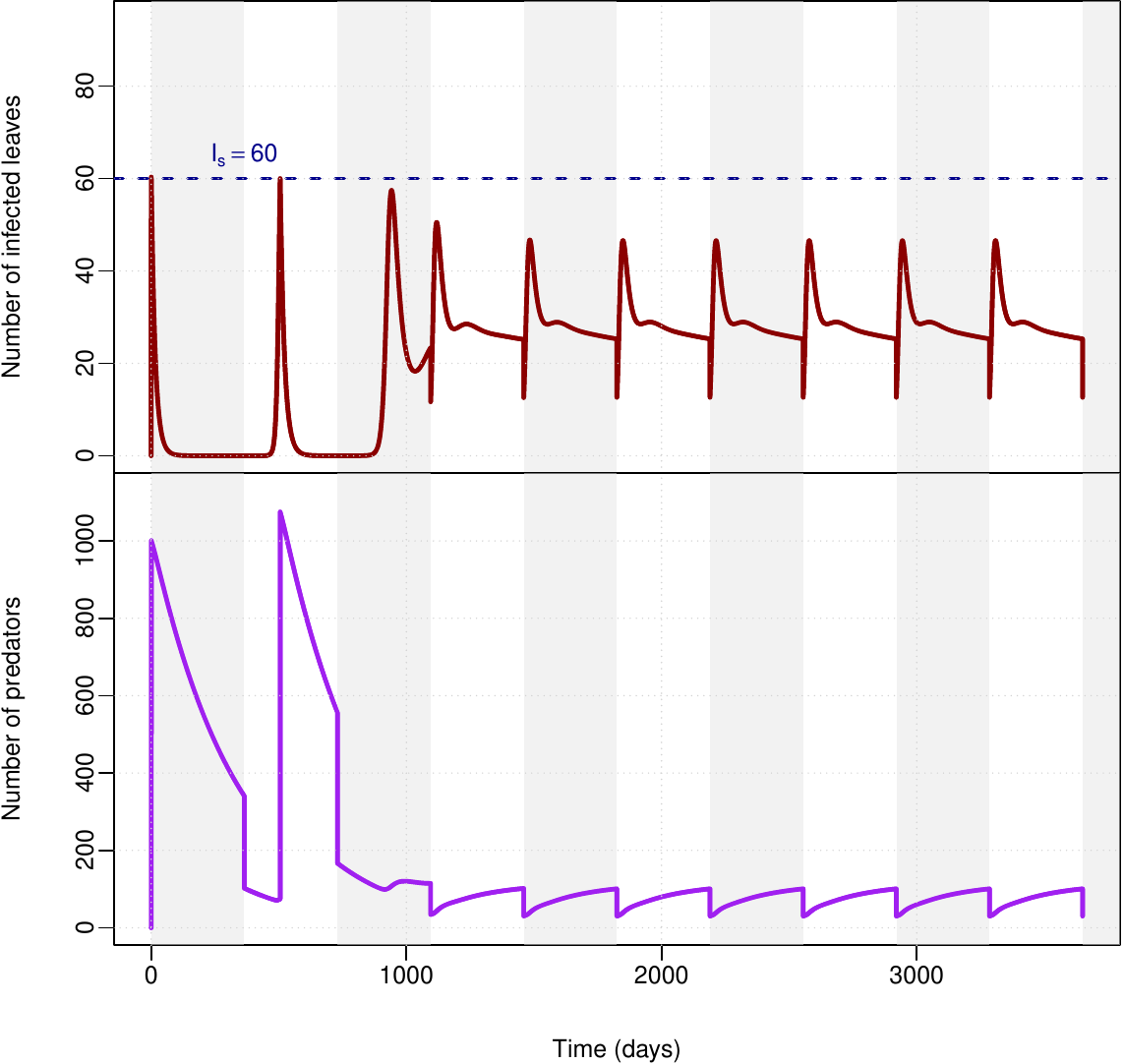}
        \label{fig:R040-Is60}
    \end{subfigure}
    \begin{subfigure}{0.47\textwidth}
        \caption{$\R_0=50$ and $I_s=60$.}
        \includegraphics[width=\textwidth]{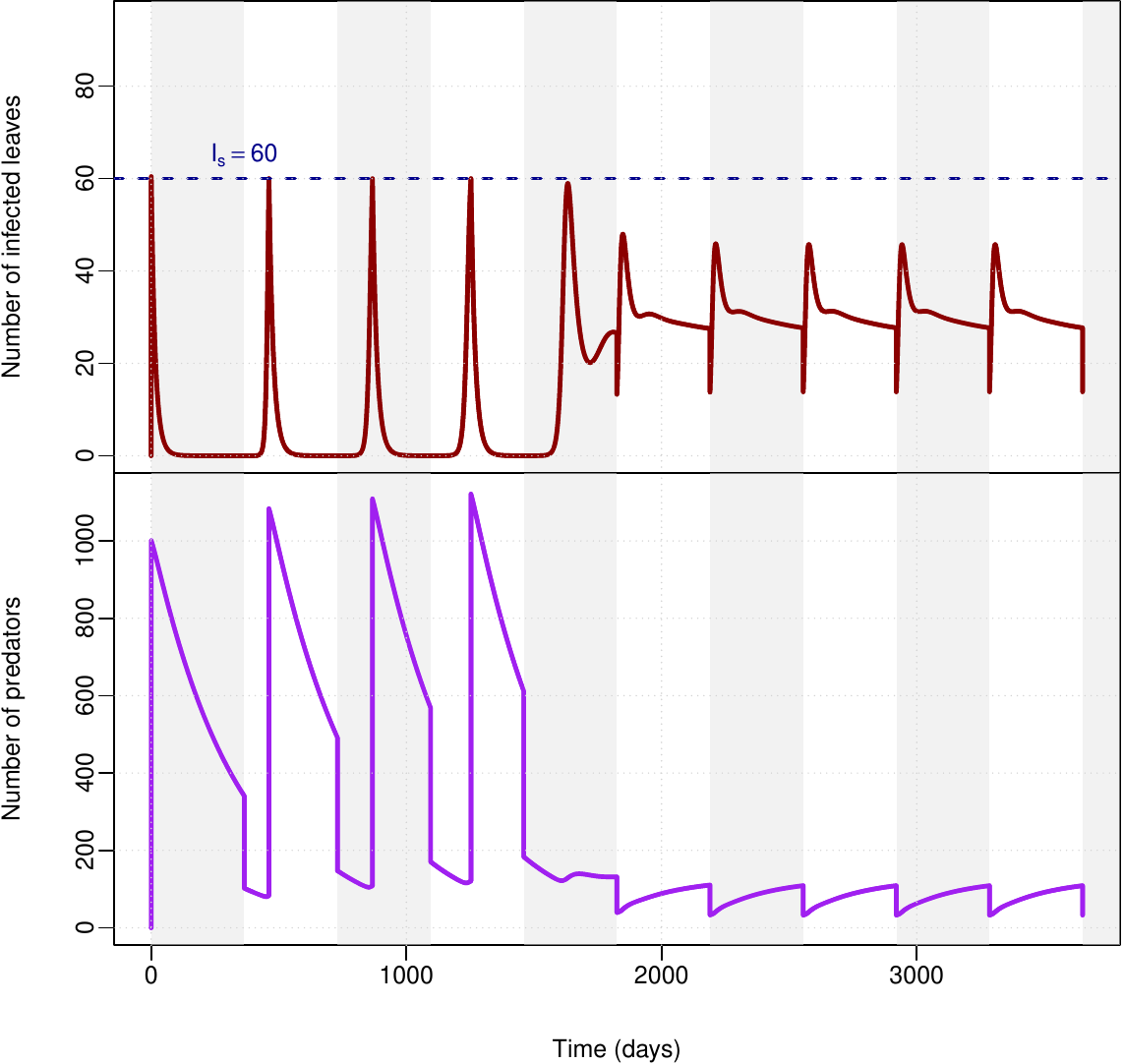}
        \label{fig:R050-Is60}
    \end{subfigure}
    \caption{Dynamics of infected leaves $I$ (top panel) and predators $P$ (bottom panel) as a function of the threshold for biocontrol $I_s$ (blue dashed lines) and $\R_0$. 
    Years shown by bands of alternating colours. 
    10-year simulations of model~\eqref{sys:imp} using initial conditions~\eqref{eq:int-cond}. 
    (a) $I_s=10$; (b) $I_s=30$; (c) $I_s=40$;   (d) $I_s=50$. 
    All parameter values as in Table~\ref{tab:model-parameters} except $\omega$ computed from $\R_0$.}
    \label{fig:sols-large-R0-Is}
\end{figure}

\paragraph{Impact of the threshold for larger values of $\mathcal{R}_0$.} 
Coffee leaf rust is not a human or animal disease and estimates for the value of $\R_0$ are scarce, so we tested a whole range of values.
A few results are shown in Figure~\ref{fig:sols-large-R0-Is}.

They illustrate several characteristics of the system. 
First, as $\R_0$ increases, the frequency of peaks when the situation is under control because of a low threshold increases.
Compare Figure~\ref{fig:R01.5-Is10} and Figure~\ref{fig:R050-Is10}. In both, $I_s=10$, but with the increase of $\R_0$ in Figure~\ref{fig:R050-Is10}, predator introductions need to become annual in order to keep the infection under control.

Increasing the threshold in turn can lead to complicated dynamics, as seen in Figure~\ref{fig:R040-Is50}.
There, the interaction between predator releases and harvest and other cultural practices leads to a dynamics that is quite complicated.

Now compare Figure~\ref{fig:R040-Is60} and Figure~\ref{fig:R050-Is60}. 
Both have $I_s=60$.
This shows an interesting characteristic: for a given threshold, the outcome is counter-intuitively better when the reproduction number is larger.
Indeed, in Figure~\ref{fig:R040-Is60} where $\R_0=40$, the situation remains under control for two years before reverting to the periodic endemic solution, whereas in Figure~\ref{fig:R050-Is60} with $\R_0=50$, it remains under control for 4 years.
Since the impact of the disease can be measured in terms of the cumulative total number of leaves infected, this means the situation with $\R_0=50$ is better.

Figure~\ref{fig:R040-Is60} and Figure~\ref{fig:R050-Is60} also illustrate the fact that thresholds should not be set in stone.
A cultivator having set their threshold at $I_s=60$, for instance, should take into account the change observed in year 3 (Figure~\ref{fig:R040-Is60}) or year 5 (Figure~\ref{fig:R050-Is60}), where rather than going to unobservable levels, the number of infected leaves rebounds during the season.
This change should trigger a lowering of the threshold $I_s$, in order to move to a situation more akin to the one in Figure~\ref{fig:R050-Is10}.

\paragraph{Measures of disease severity.} 
In order to get a better sense of the impact of biocontrol measures, we consider the the total number of newly infected leaves over the period $[0,T]$,
\[
I_{\text{total}}(T) = \int_0^T \omega\nu S(s)U(s)\ ds,
\]
i.e., the total incidence.

Figure~\ref{fig:heatmaps} shows several heat map plots, which help better understand the earlier discussion.
These plots were obtained by considering, for each point, the solution of the IDE~\eqref{sys:imp-control} for a period of 10 years.

First, note that Figure~\ref{fig:heatmaps-u0-IS-R01.5} and Figure~\ref{fig:heatmaps-u0-IS-R050} show that the situation is more nuanced when the basic reproduction number $\R_0$ is larger, confirming our earlier discussion about Figure~\ref{fig:R040-Is60} and Figure~\ref{fig:R050-Is60}.
Indeed, we observe there that when $\R_0=1.5$, the situation changes abruptly from a low to a high overall infection, whereas the change is more progressive when $\R_0=50$.
In both cases, the initial number $U_0$ does not fundamentally change the outcome, even less so when $\R_0$ is high.

In Figure~\ref{fig:heatmaps-R0-IS-muP0.003} and Figure~\ref{fig:heatmaps-R0-IS-muP0.01}, we show the change in total infection as a function of the basic reproduction number $\R_0$ and the threshold $I_s$, for two different values of the predator death rate $\mu_P$.
The situation is quite similar in both cases, except that when the death rate of predators is lower, a slightly lower overall burden of infection is reached, supporting the findings in \cite{djuikem2023impulsive}. 
These graphs also illustrate the complexity of the situation: there are regions with medium range values of $\R_0$ from around 10 to around 20 that are much more tolerant to a higher value of $I_s$ than regions with both higher and lower values of $\R_0$; this is observed in both Figure~\ref{fig:heatmaps-R0-IS-muP0.003} and Figure~\ref{fig:heatmaps-R0-IS-muP0.01}, albeit in a more pronounced manner when the death rate of predators is smaller (Figure~\ref{fig:heatmaps-R0-IS-muP0.003}).

Figure~\ref{fig:heatmaps-muP-IS-u0200} again confirm the complexity of the situation, with the change between the regions where control is maintained and where it first degrades being quite complicated (horizontal band between roughly $I_s=50$ and $I_s=80$), then abrupt in the switch from some control to no control at all at around $I_s=80$.
Figure~\ref{fig:heatmaps-muP-IS-u0200} uses a set value of $\R_0=5$ and and initial spore count $U_0=200$.
The situation is similar (not shown) for different values of $U_0$, with the threshold between partially controlled to uncontrolled disease decreasing as $U_0$ increases.

We take this opportunity to dig further into the abrupt switch observed in all figures, by considering Figure~\ref{fig:heatmaps-muP-IS-u0200-zoom}, which zooms into the region where the switch occurs between partially controlled to uncontrolled disease in Figure~\ref{fig:heatmaps-muP-IS-u0200}.
All heatmaps shown in Figure~\ref{fig:heatmaps} were obtained by considering integer values of $I_s$.
What we deduce from Figure~\ref{fig:heatmaps-muP-IS-u0200-zoom} is that for the parameters used there, there is some value $I_\text{max}$ of $I$ in $(80,81)$ that is the maximum of $I(t)$ along the solution over the ten year period considered here. 
Setting $I_s=80$ means that each solution (for the different values of $\mu_P$ shown in Figure~\ref{fig:heatmaps-muP-IS-u0200-zoom}) hits $I_s=80$ before reaching $I_\text{max}$, triggering a predator introduction. 
On the other hand, setting $I_s=81>I_\text{max}$ means that solutions never reach the threshold $I_s$.
In terms of the dynamics, this is the situation observed in Figure~\ref{fig:R01.5-Is70}, where $I_\text{max}$ must take a value slightly larger than 60.
This also illustrates that the difference between no introduction and even a single introduction is quite stark: in Figures~\ref{fig:heatmaps-muP-IS-u0200} and \ref{fig:heatmaps-muP-IS-u0200-zoom}, values in the yellow region are all equal to 10,984 new infected leaves over a period of ten years, while values for $I_s=80$ range from 625 to 9,152 new infected leaves during the same period.

\section{CTMC and branching process models}
\label{sec:CTMC-MBPA}

The non-intuitive dynamics we discussed, where biocontrol effectiveness does not always align with expectations, particularly for different threshold values $I_s$, leads to an important question: if the farmer chooses not to implement biocontrol measures, will the disease continue to spread as predicted by the IDE?
Since the germination rate $\omega$ was the parameter used earlier to calculate different \(\mathcal{R}_0\), it is also interesting to determine in which conditions the infection is most likely to disappear naturally in the plantation.
Finally, resurgences as observed in Figure~\ref{fig:R01.5-Is10} and Figure~\ref{fig:R01.5-Is50}, where there are several years between disease outbreaks, suggest that perhaps using an ODE leads to an atto-fox problem \cite{mollison1991dependence} and that a real extinction should have happened that is precluded by the nature of the system.

To explore this further, we use a continuous-time Markov chain (CTMC) model as well as a branching process approximation of the CTMC to investigate extinction probabilities and stochastic dynamics.
This allows to consider the scenarios of CLR extinction in the absence of biocontrol. 
The idea is to evaluate the percentage of plantations or cases where the epidemic eventually dies out spontaneously without requiring any control intervention. 
This helps determine whether, in certain circumstances, the decision by some farmers not to apply treatment in their plantations might be justified.


\subsection{Continuous time Markov chain model}

The ODE model is limit of a CTMC~\cite{Kurtz1970}. We consider here CTMC model related to the deterministic model~\eqref{sys:ode-model}, i.e., the ODE in the non-periodic and uncontrolled stage.  
Unlike the ODE or IDEs, the CTMC model tracks discrete counts of individuals, allowing for true disease extinction.

The CTMC model related to the deterministic model~\eqref{sys:ode-model} has state
\begin{subequations}\label{sys:CTMC_general}
    \begin{equation}\label{sys:CTMC_general_Xt}
        \mathbf{X}_t=
        \left( S(t),I(t),U(t),P(t)\right),\quad t\in(0,T]
\end{equation}
where each element of the vector is a discrete random variables that take values in $\mathbb{N}$ and where the time between events is exponentially distributed \cite{allen2010introduction}. Following the classical formulation of CTMC~\cite{allen2010introduction, kurtz1972relationships}, $\mathbf{X}_t$ is  characterised by transition probabilities from state $\mathbf{k}$ to state $\mathbf{j}$,
\begin{equation}\label{sys:CTMC_general_proba}
\mathbb{P}(\mathbf{X}(t+\Delta t)=\mathbf{j} \mid \mathbf{X}(t)=\mathbf{k})=\sigma(\mathbf{k}, \mathbf{j}) ,    
\end{equation}
\end{subequations}
with rates $\sigma(\mathbf{k}, \mathbf{j})$ given in Table~\ref{tab:prob-CTMC}.

\begin{table}[H]
\caption{Reaction rates used to determine transition probabilities for the general CTMC model \eqref{sys:CTMC_general}.}
\centering
\begin{tabular}{lll}
\hline Event & Transition  $\mathbf{k} \rightarrow \mathbf{j}$ & Rate  $\sigma(\mathbf{k}, \mathbf{j})$ \\
\hline
Birth of $S$ & $S\rightarrow S+1$ & $\Lambda$ \\
Natural death of $S$ & $ S\rightarrow S-1$ & $\mu S$ \\
Natural death of $I$ & $I\rightarrow I-1$ & $\mu I$ \\
Infection of $S$ by $U$ & $S \rightarrow S-1, I \rightarrow I+1$ & $\omega \nu U S$ \\
Death of $I$ due to CLR & $I \rightarrow I-1$ & $d I$ \\

Production of $U$ & $ U \rightarrow U+1$ & $\gamma I$ \\
Natural death of $U$ & $U\rightarrow U-1$ & $\mu_U U$ \\
Death of $U$ due to germination & $U\rightarrow U-1$ & $\nu U S$ \\
Consumption of $U$ by $P$ & $U \rightarrow U-1$ & $\delta U P$ \\
Transformation of $U$ to $P$ & $P \rightarrow P+1$ & $ \eta \delta UP$ \\
Natural death of $P$ & $P \rightarrow P-1$ & $\mu P$ \\
\hline
\end{tabular}  
\label{tab:prob-CTMC}
\end{table}

\subsection{Multitype branching process approximation}

Following the definition of multitype branching process given in \cite{athreya1972branching, Haccou_Jagers_Vatutin_2005, Harris1963, kurtz1972relationships}, denote \(Y(t)=(Y_I(t), Y_U(t))\) the vector random variable consisting of $2$ infected types, where  $Y_i(t)$ is the discrete random variable for the number of individuals in infectious type \( k=\{I,U\} \) at time \( t \in [0, \infty) \), \( Y_k(t) \in \mathbb{N}_0 = \{0, 1, 2, \ldots\} \).

For this branching process, define the probability generating function (p.g.f.) for type \( k \) at time \( t \) as
\[
F_k(t, \bz) = \mathbb{E}\left(\bz^{Y(t)} | Y(0) = e_k\right),
\]
where \( e_k \) is the \( k \)-th unit vector, \( \bz = (z_I,z_U) \in [0, 1]^2 \), and the notation \( \bz^{Y(t)} = \left(z_I^{Y_I(t)}, z_U^{Y_U(t)} \right)\). Let 
\[
F(t, \bz) = (F_I(t, \bz), F_U(t, \bz))
\]
denote the p.g.f. for the entire process.

The p.g.f. \( F_k(t,\bz) \) for type \(k\) of the entire stochastic process is a solution of the backward Kolmogorov differential equation,
\[
\frac{\partial}{\partial t} F_k(t,\bz) = \sigma_k [ f_k(F(t,\bz)) - F_k(t,\bz)], \quad k = 1,2.
\]
with initial conditions \( F_k(0,\bz) = z_k \), where \( \sigma_k \) is the rate parameter for the exponentially distributed lifetime of type  \(k\) \cite{athreya1972branching, Haccou_Jagers_Vatutin_2005, Harris1963,  kurtz1972relationships}, where $f_k$ is the offspring p.g.f. for type \(k\), such that, given \( Y(0) = e_k \), this offspring is defined as \( f_k: [0,1]^2 \to [0,1] \),
\[
f_k(z_I,z_U) = \sum_{k_U=0}^{\infty}\sum_{k_I=0}^{\infty} P_k(k_I, k_U) z_I^{k_I}z_U^{k_U},
\]
with \(P_k(k_I,k_U)\) the probability that an individual of type \(k\) gives birth to \(k_I\) individuals of type \(z_I\), \(k_U\) individuals of type \(z_U\). 
Using the transition rates in Table~\ref{tab:prob-CTMC}, the offspring p.g.f is  defined as
\begin{equation}\label{eq:pgf}
    f(z)=(f_I(z),f_U(z)),
\end{equation}
where
\begin{subequations}\label{sys:offs-pgf}
    \begin{align}
    f_I(z)& =\frac{\gamma z_U z_I+\mu+d}{\gamma+\mu+d}\label{sys:pgf_I}\\
    f_U(z)& =\frac{\omega\nu S^0 z_I+\nu S^0+\mu_U}{\omega\nu S^0+\nu S^0+\mu_U} \label{sys:pgf_U}.
\end{align}
\end{subequations}
The Jacobian matrix of \eqref{sys:offs-pgf} takes the form
\[
J_f(z)=\begin{bmatrix}
\dfrac{\gamma z_{U}}{d + \gamma + \mu} & \dfrac{\gamma z_{I}}{d + \gamma + \mu}\\
\dfrac{S_{0} \nu \omega}{S_{0} \nu \omega + S_{0} \nu + \mu_{U}} & 0
\end{bmatrix}.
\]

\begin{theorem}\label{th:exists-fp-importer}
    For the initial infected leaves \( I(0) = i_0 \) and spores \( U(0) = u_0 \), the probability of extinction of the CLR is given by:
    \begin{subequations}
    \label{eq:prob-ext-CLR}
    \begin{align}
        \mathbb{P}_{\text{ext}} &= \left(q_I\right)^{i_0} \times \left(q_U\right)^{u_0}, \\
        \mathbb{P}_{\text{outbreak}}&= 1 - \mathbb{P}_{\text{ext}},
    \end{align}    
    \end{subequations}
    where 
    \begin{equation*}\label{eq:def-q}
        q := (q_I, q_U)
    \end{equation*}
    is a fixed point on \( [0,1]^{2} \) of the offspring p.g.f~\eqref{eq:pgf}. The following alternative holds:
    \begin{itemize}
        \item If \( \mathcal{R}_0 \leq 1 \), then \( q = \mathbf{1} \), i.e., \( \mathbb{P}_{\text{ext}} = 1 \), where \( \mathbf{1} \) is the unit column vector of size \( 2 \);
        \item If \( \mathcal{R}_0 > 1 \), then in addition to \( q = \mathbf{1} \), there is a unique vector \( \mathbf{0} < q < \mathbf{1} \) such that \( J_f(q) = q \).
    \end{itemize}
\end{theorem}

\begin{proof}
We make two observations.
\textit{(i)} 
The matrix $J_f(q)$ is monotone since \(\forall u \in[0,1)^{2}\), \( J_f(q) \geq 0 \) implies that \( q \geq 0 \).
As a consequence, the multitype branching processes are not singular \cite[Theorem 2.3 p. 113]{berman1979nonnegative}.
\textit{(ii)}
The matrix of first moments \( \mathbb{M} = D\bF(\mathbf{1}) \) is primitive, since \(\mathbb{M}^2\) is positive \cite{berman1979nonnegative}.

From (i) and (ii), we conclude that the branching process is positive and regular. As a consequence, applying the Threshold Theorem \cite{ALLEN201399} together with \cite[Theorem 7.1 p. 41]{Harris1963} concludes the proof of Theorem~\ref{th:exists-fp-importer}.
\end{proof}

The fixed point of \eqref{sys:offs-pgf} is $(f_I(q_I,q_U), f_U(q_I,q_U))=(q_I,q_U)$; therefore, for \eqref{sys:pgf_I}, one has
\[
q_I=\frac{(\omega\nu S^0+\nu S^0+\mu_U)(\mu+d)}{\gamma\omega\nu S^0} \text{ and } q_U=   \frac{\mu+d}{\gamma}+\frac{\nu S^0+\mu_U}{\omega\nu S^0+\nu S^0+\mu_U}. 
\]
Replacing the above values, it follows that the probability of extinction is given by
\begin{equation*}
   \mathbb{P}_{\text{ext}}=\left\{ 
    \begin{aligned}
     &  \left(q_I \right)^{i_0} \left(q_U\right)^{u_0}, \, &\mathcal{R}_0 >1\\
     &  1 , \; &\mathcal{R}_0 \leq 1\
    \end{aligned}
    \right.
\end{equation*}
and the probability of an outbreak is 
\begin{equation}\label{eq:proba-outbreak}
  \mathbb{P}_{\text{outbreak}}=1- \mathbb{P}_{\text{ext}}.
\end{equation}

\subsection{Numerical simulation}

\begin{figure}[htbp]
    \centering
    \includegraphics[width=0.5\linewidth]{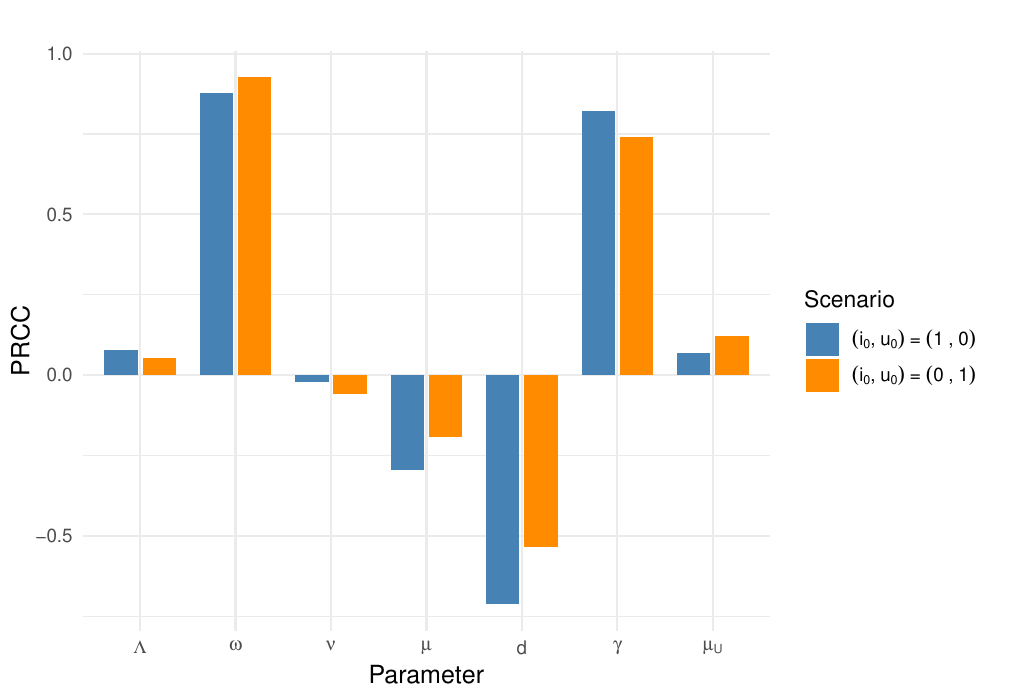}
    \caption{Partial rank correlation coefficients (PRCC) for the sensitivity of the probability of an outbreak to changes in parameters. Two initial conditions are shown, one where the infection starts with one infected leaf \( (i_0, u_0) = (1, 0) \) and the other with one spore \( (i_0, u_0) = (0, 1) \).}
    \label{fig:proba-outbreak}
\end{figure}

To understand the impact of parameters on the probability \eqref{eq:proba-outbreak} of an outbreak, we conduct a sensitivity analysis. 
Figure~\ref{fig:proba-outbreak} shows PRCC values of parameters with respect to the probability of an outbreak. 
Parameters such as the germination rate \(\omega\) and the spore production rate \(\gamma\) exhibit strong positive effects on the probability of outbreak, while the disease-induced death rate \(d\) and the natural death rate \(\mu\) have negative effects. 
Moreover, the impact remains consistent across both initial conditions. 
Comparing Figures~\ref{fig:prcc-analysis} and~\ref{fig:proba-outbreak}, we observe that the parameters \( \Lambda \), \( \omega \), \( \nu \), \( \mu \), \( d \), \( \gamma \) and \( \mu_U \) have relatively similar impacts on both the basic reproduction number \( \mathcal{R}_0 \) and the probability \(\mathbb{P}_{\text{outbreak}}\) of an outbreak.

\begin{figure}[htbp]
    \centering
    \begin{subfigure}{0.45\textwidth}
        \includegraphics[width=1\linewidth]{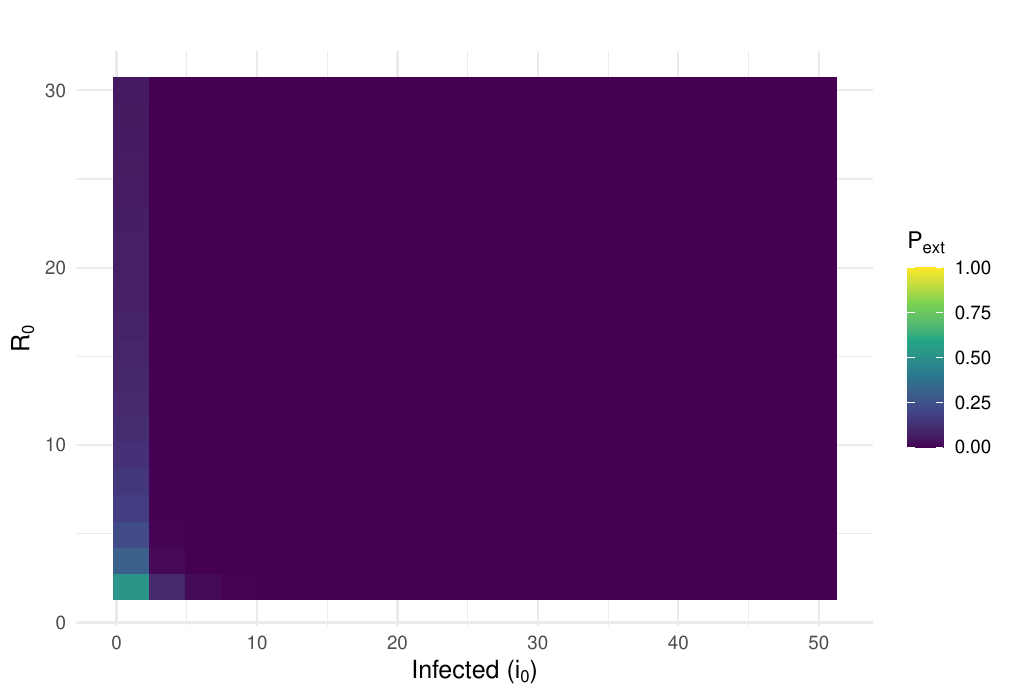}        
        \caption{$(i_0,u_0)=(1,0)$.}
        \label{fig:heatmap-pro-ext-i0}
    \end{subfigure}
    \begin{subfigure}{0.45\textwidth}
        \includegraphics[width=1\textwidth]{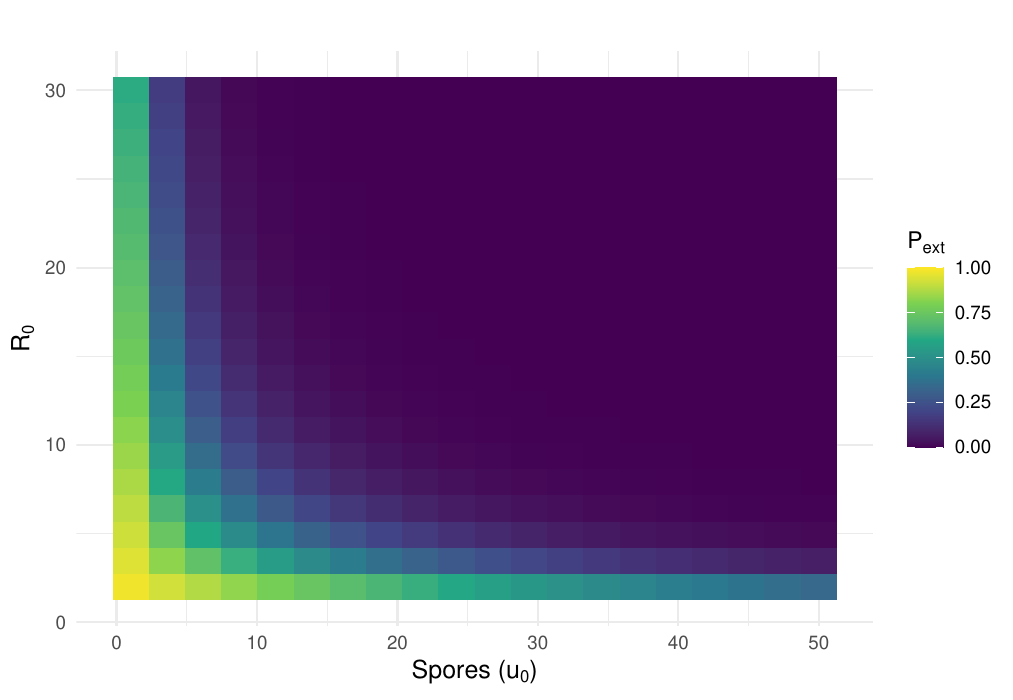}
        \caption{$(i_0,u_0)=(0,1)$.}
        \label{fig:heatmap-pro-ext-u0}
    \end{subfigure}
    \caption{Heatmap of the probability of CLR extinction $\mathbb{P}_{\text{ext}}$ as a function of the basic reproduction number $\mathcal{R}_0$ for initial conditions for the infection with (a) one infected leaf \( (i_0, u_0) = (1, 0) \) and (b) one spore \( (i_0, u_0) = (0, 1) \).}
    \label{fig:heatmap-pro-ext}
\end{figure}

Figure~\ref{fig:heatmap-pro-ext} presents two heatmaps illustrating the probability $\mathbb{P}_{\text{ext}}$ of CLR extinction as a function of the basic reproduction number $\mathcal{R}_0$ (ranging from 0 to 35) and two initial conditions ($i_0, u_0$).
Figure~\ref{fig:heatmap-pro-ext-i0}  shows $\mathbb{P}_{\text{ext}}$ as a function of the initial number of infected individuals $i_0$  for different values of $\mathcal{R}_0$, assuming no initial spores ($u_0 = 0$). 
The probability of extinction is $\mathbb{P}_{\text{ext}} = 0$, in purple for almost all values indicating a more persistent infection. 
Except when $i_0=2$ and $\mathcal{R}_0$ is below 5, the probability of extinction is small ($\mathbb{P}_{\text{ext}} \approx 0.4$, in green), meaning CLR extinction.
Figure~\ref{fig:heatmap-pro-ext-u0} illustrates the probability of disease extinction, $\mathbb{P}_{\text{ext}}$, as a function of the initial number of spores, $u_0$ for various values of $\mathcal{R}_0$, assuming no initial infections ($i_0 = 0$). When $u_0 < 10$ and $\mathcal{R}_0 < 10$, disease extinction is highly probable ($\mathbb{P}_{\text{ext}} \approx 1$, shown in yellow). However, for larger $u_0$ and sufficiently high $\mathcal{R}_0$, the probability of disease extinction decreases significantly toward zero. This indicates that a high  initial spore ($u_0>20)$ concentration favours disease persistence, even in the absence of initial infected leaves.

In Figure~\ref{fig:heatmap-pro-ext}, we observe that the scenario in which the system initially starts with spores is the one where extinctions are most easily observed. 
This is somewhat validating of the model, since in practice, infection in a new plantation necessarily occurs through spores.

These results highlight that, even in the absence of external control measures, such as biological control with predators, CLR can still go extinct naturally due to stochastic fluctuations, particularly when the initial spore count is small (below 30) and $\mathcal{R}_0$ is below 5.
However, when the initial spore quantity is high or when $\mathcal{R}_0$ is large, the disease is more likely to persist, emphasizing the need for stronger management strategies.

\section{Discussion}
\label{sec:Discussion}
This study first analyses the dynamics of deterministic models of coffee leaf rust under threshold-based impulsive biocontrol strategies.
The mathematical analysis does not extend to the impulsive model with state-based control measures.
Indeed, the Floquet theory used to conduct the stability analysis of the model with annual impulses does not work in the case where predators are released in a state-dependent non-periodic manner, since this irregularity leads to a new monodromy matrix at the end of each year.

A computational analysis of the deterministic models highlights the importance of well-timed interventions, particularly for smallholder farmers who often rely on reactive treatments.
This analysis also underscores the complex link between parameter space and model dynamics, with potential outcomes of an introduction strategy varying greatly depending on the threshold chosen; see, e.g., Figure~\ref{fig:heatmaps}.
Altogether, though, the computational analysis makes it clear that a low threshold for introduction is generally better and that setting too high a threshold likely leads to completely missing on the opportunity to control the infection.

Variations observed for the deterministic impulsive differential equations model with biocontrol motivated the use of a stochastic framework to compute the probability of coffee leaf rust extinction, indicating that CLR may naturally disappear when initial infection levels are low. 
This could explain why some farmers delay intervention: they have first-hand experience of ``failed invasions'' and therefore believe this will happen again.
However, the CTMC also shows that when the reproduction number or the initial number of spores are large, the probability that the disease will go extinct becomes small, highlighting the need for control measures.

Note that the continuous time Markov chain model was only applied to the first year of spread and did not incorporate impulsive events.
Future explorations of the models should incorporate these events, in order in particular to clarify whether long term resurgences as seen in the deterministic IDE models are an artefact of the modelling paradigm or are actually possible.

Finally, another aspect not taken into account in our model is the diffusive nature of CLR spore spread.
Such spatial spread is determinant in plant diseases, since the latter are essentially immobile.
Threshold-based impulsive biocontrol would be a natural and interesting extension of the work of the present work and \cite{djuikem2021modelling}, which analysed a CLR model set in a partial differential equation framework accounting for the spatial dynamics of spores.

\section*{Acknowledgements}
JA acknowledges support from NSERC through the Discovery Grants program. This work was partially supported by a 2024/2025 MMI Research Accelerator Award and we gratefully acknowledge the Maud Menten Institute (PRN2).

\appendix

\section{Proofs}
\label{sec:appen-proof-theo-gas}

\begin{proof}[Proof of Lemma~\ref{lm:LAS-DFE}]
Let us prove the result in two steps.

\noindent\textit{Step 1 --}
Let $y(t)=(I(t),U(t))$. 
Consider the subsystem of \eqref{sys:ode-model} defined as follows
\begin{equation}\label{sys:sub-stability-IU-ODE}
        \dot y(t) = F_c(y(t)),
\end{equation}
where 
\[
F_c(y)=\begin{pmatrix} 
\omega \nu S U - (\mu + d) I \\ 
\omega \gamma I - \nu U S - \mu_U U - \delta U
\end{pmatrix}.
\]
Consider the Lyapunov function candidate $V: \mathbb{R}^2_+ \to [0, \infty) $  defined by 
\[
V(I,U)=I+aU,
\]
where $a$ is a positive constant to be chosen later. 
Let $\mathcal{G}=\{(I,U)\in \mathbb{R}_+^2, I\leq \Gamma_N, U \leq \Gamma_U\}$; then the following properties hold.
\begin{itemize}
    \item[(i)]   For  $y\in \mathcal{G}_1$, $V(y)>0$ and   $V(0,0)=0$.
    \item [(ii)] Let us prove that $ V'(y)F_c(y)\leq 0$. We have
    \begin{align*}
        V'(y)F_c(y)&= \omega \nu S U - (\mu + d) I+a\left( \gamma I - \nu U S - \mu_U U - \delta U\right)\\
        & =  \left(\omega \nu S-a(\nu  S + \mu_U  )\right)U +(a \gamma -(\mu+d))I - a\delta U.
    \end{align*}
    If $a\gamma-(\mu+d)=0$, i.e., $a=(\mu+d)/\gamma$, then
    \[
    V'(y)F_c(y)= \left(\omega \nu S-\frac{\mu+d}{\gamma}(\nu  S + \mu_U  )\right)U - a\delta U.
    \]
    For all $S\in [0, S^0] $, define $g(S)=\omega \nu S-(\mu+d)(\nu  S + \mu_U)/{\gamma}$. Then 
    \[
    g(0)=0 \text{ and } g(S^0)=\omega \nu S^0-\frac{\mu+d}{\gamma}(\nu  S^0 + \mu_U).
    \]

    If $\mathcal{R}_0<1$, with $\R_0$ defined by \eqref{eq:R0}, then the function $g$ is decreasing from $0$ to $g(S^0)<0$. 
    So if $\mathcal{R}_0<1$, \(g(S)<0\), and as a consequence, $ V'(y)F_c(y) < 0$. 
    \item[(iii)] Finally one has $V(y)\to\infty$ as $||y|| \to\infty$.
\end{itemize}

Then it is easy to see that the largest invariant set defined over $\{y\in \mathcal{G}_1, V'(y)=0\}$ is the singleton $\{(0,0)\}$. 
Therefore, using the LaSalle invariance principle \cite{la1976stability}, we deduce from conditions (i)--(iii) above that the solution $y_e=(0,0)$ is GAS under the condition $\mathcal{R}_0 < 1$ for subsystem~\eqref{sys:sub-stability-IU-ODE}.

\noindent\textit{Step 2 --}
Let now $x(t)=(S(t),P(t))$.
Substituting the solution $U=0$ into~\eqref{sys:ode-model}, we obtain the following subsystem 
\begin{equation}\label{sys:sub-stability-SP-ODE}
    \dot x(t) = F_c(x(t)), \quad
\end{equation}
where
\[
    F_c(x)=\begin{pmatrix} 
        \omega \Lambda - \mu S \\ 
        \omega \mu_P P
    \end{pmatrix}.
\]
System~\eqref{sys:sub-stability-SP-ODE}  has the unique equilibrium $(S^0(t),0)$, which is GAS. 
The solutions of \eqref{sys:ode-model} are inside $\mathcal{G}$, then $(S(t),P(t)) \to (S^0(t),0)$.

Using steps 1 and 2, one concludes that the DFE $\bE^0(t)$ is globally asymptotically stable for \eqref{sys:ode-model}. This concludes the proof.
\end{proof}

\begin{proof}[Proof of Theorem~\ref{thm:gloal-sta-PDFS}]
The proof follows that of Lemma~\ref{lm:LAS-DFE} with some adaptation to the IDE case.
As before, we proceed in two steps.

\noindent\textit{Step 1 --}
Additionally to $y(t)=(I(t),U(t))$, define $\Delta y(t)=y(t^+)-y(t)$ for any $t\geq 0$ and consider the subsystem of \eqref{sys:imp} defined as follows
\begin{equation}\label{sys:sub-stability-IU}
    \begin{aligned}
        \dot y(t) &= F_c(y(t)),\quad t\neq n\tau \\
        \Delta y(t) &= F_d(y(t)),\quad  t=n\tau,
    \end{aligned}
\end{equation}
where 
\[
F_c(y)=\begin{pmatrix} 
\omega \nu S U - (\mu + d) I \\ 
\omega \gamma I - \nu U S - \mu_U U - \delta U
\end{pmatrix}
\text{ and } 
F_d(y)=\begin{pmatrix} 
(\varphi_I-1)I\\
( \varphi_U-1)U\end{pmatrix}.
\]
Consider the same Lyapunov function candidate $V(I,U)=I+aU$ as in the proofof Lemma~\ref{lm:LAS-DFE} and the same set $\mathcal{G}_1=\{(I,U)\in \mathbb{R}_+^2, I\leq \Gamma_N, U \leq \Gamma_U\}$. 
Then properties (i)--(iii) in Step 1 of the proof of Lemma~\ref{lm:LAS-DFE} hold here too. 
Additionally, the following property holds true.
\begin{itemize}
    \item [(ii$^\prime$)] We have 
    \begin{align*}
        V(y+F_d(y))&=I+(\varphi_I-1)I+a(U+( \varphi_U-1)U)\\
        &= \varphi_I I + a \varphi_U U\\
        &< I+a U, \text{ using the fact that } \varphi_I, \varphi_U <1,\\
        &< V(y).
    \end{align*}
\end{itemize}

Clearly, the largest invariant set defined over $\{y\in \mathcal{G}_1, V'(y)=0\}$ is the singleton $\{(0,0)\}$. 
Therefore, using the LaSalle-Krasowski invariant principle adapted to IDE \cite[Theorem 4.2]{haddad2006impulsive}, we deduce from conditions (i)--(iii) and (ii$^\prime$) that the solution $y_e=(0,0)$ is GAS under the condition $\mathcal{R}_0 < 1$ for subsystem~\eqref{sys:sub-stability-IU}.

\noindent\textit{Step 2 --} 
Let now $x(t)=(S(t),P(t))$.
If we substitute the solution $U=0$ into~\eqref{sys:imp}, we obtain the following subsystem 
         \begin{equation}\label{sys:sub-stability-SP}
            \begin{aligned}
                \dot x(t) &= F_c(x(t)), \quad t\neq n\tau \\
                \Delta x(t) &= F_d(x(t)), \quad t=n\tau,
            \end{aligned}
        \end{equation}
        where
        \[
        F_c(x)=\begin{pmatrix} 
        \omega \Lambda - \mu S \\ 
        \omega \mu_P P
        \end{pmatrix} 
        \text{ and }
        F_d(x)=\begin{pmatrix} 
        (\varphi_S-1)S\\
        ( \varphi_P-1)P\end{pmatrix}.
        \]
        System~\eqref{sys:sub-stability-SP}  has the unique periodic solution $(S^T(t),0)$, which is GAS. 
        The solution of \eqref{sys:imp} are inside $\mathcal{G}$, then $(S(t),P(t)) \to (S^T(t),0)$.

Using steps 1 and 2, one concludes that the PDFS $\bE^T(t)$ is globally asymptotically stable for \eqref{sys:imp}. This concludes the proof.
\end{proof}

\bibliographystyle{plain}
\bibliography{references}

\begin{thebibliography}{10}

\bibitem{allen2010introduction}
L.J.S. Allen.
\newblock {\em An Introduction to Stochastic Processes with Applications to
  Biology}.
\newblock Chapman and Hall/CRC, an imprint of Taylor and Francis, second
  edition, 2010.

\bibitem{ALLEN201399}
L.J.S. Allen and P.~{van den Driessche}.
\newblock Relations between deterministic and stochastic thresholds for disease
  extinction in continuous- and discrete-time infectious disease models.
\newblock {\em Mathematical Biosciences}, 243(1):99--108, May 2013.

\bibitem{athreya1972branching}
K.B. Athreya and P.E. Ney.
\newblock {\em Branching Processes}.
\newblock Springer, 1972.

\bibitem{avelino2015impact}
J.~Avelino, M.~Cristancho, S.~Georgiou, P.~Imbach, L.~Aguilar, G.~Bornemann,
  and C.~Morales.
\newblock The coffee rust crises in {C}olombia and {C}entral {A}merica
  (2008–2013): impacts, plausible causes and proposed solutions.
\newblock {\em Food Security}, 7(2):303--321, 2015.

\bibitem{berman1979nonnegative}
A.~Berman and R.J. Plemmons.
\newblock {\em Nonnegative Matrices in the Mathematical Sciences}.
\newblock New York: Academic Press, 1979.

\bibitem{Bock1962}
K.R. Bock.
\newblock Dispersal of uredospores of \emph{{H}emileia vastatrix} under field
  conditions.
\newblock {\em Transactions of the British Mycological Society}, 45(1):63--74,
  1962.

\bibitem{brent2003fungicide}
K.J. Brent and D.W. Hollomon.
\newblock Fungicide resistance: the assessment of risk.
\newblock {\em FRAC Monograph}, 2:1--49, 2003.

\bibitem{DABA2022e11892}
G.~Daba, G.~Berecha, B.~Lievens, K.~Hundera, K.~Helsen, and O.~Honnay.
\newblock Contrasting coffee leaf rust epidemics between forest coffee and
  semi-forest coffee agroforestry systems in {SW}-{E}thiopia.
\newblock {\em Heliyon}, 8(12):e11892, 2022.

\bibitem{daivasikamani2009biocontrol}
S.~Daivasikamani and Rajanaika.
\newblock Biological control of coffee leaf rust pathogen, \textit{{H}emileia
  vastatrix} berkeley and broome using bacillus subtilis and pseudomonas
  fluorescens.
\newblock {\em Journal of Biopesticides}, 2(1):94–98, 2009.

\bibitem{agronomy11091865}
M.L.V. de~Resende, E.A. Pozza, T.~Reichel, and D.M.S. Botelho.
\newblock Strategies for coffee leaf rust management in organic crop systems.
\newblock {\em Agronomy}, 11(9), 2021.

\bibitem{djuikem2023impulsive}
C.~Djuikem, F.~Grognard, and S.~Touzeau.
\newblock Impulsive modelling of rust dynamics and predator releases for
  biocontrol.
\newblock {\em Mathematical Biosciences}, 356:108968, 2023.

\bibitem{djuikem2021modelling}
C.~Djuikem, F.~Grognard, R.~T. Wafo, S.~Touzeau, and S.~Bowong.
\newblock Modelling coffee leaf rust dynamics to control its spread.
\newblock {\em Mathematical Modelling of Natural Phenomena}, 16:1--25, 2021.

\bibitem{Djuikem2025busseola}
C.~Djuikem and J.~Tchouanti.
\newblock Dynamics and control of maize infection by \emph{Busseola fusca}:
  multi-seasonal modeling and biocontrol strategies.
\newblock {\em arXiv preprint arXiv:2503.16615}, 2025.

\bibitem{DUPRE2022105918}
S.I. Dupre, C.A. Harvey, and M.B. Holland.
\newblock The impact of coffee leaf rust on migration by smallholder coffee
  farmers in guatemala.
\newblock {\em World Development}, 156:105918, 2022.

\bibitem{Haccou_Jagers_Vatutin_2005}
P.~Haccou, P.~Jagers, and V.A. Vatutin.
\newblock {\em Branching Processes: Variation, Growth, and Extinction of
  Populations}.
\newblock Cambridge University Press, 2005.

\bibitem{haddad2006impulsive}
W.M. Haddad, V.~Chellaboina, and S.G. Nersesov.
\newblock {\em Impulsive and hybrid dynamical systems: stability,
  dissipativity, and control}.
\newblock Princeton University Press, 2006.

\bibitem{HAJIANFOROOSHANI2023105099}
Z.~Hajian-Forooshani, I.~Perfecto, and J.~Vandermeer.
\newblock Novel community assembly and the control of a fungal pathogen in
  coffee agroecosystems.
\newblock {\em Biological Control}, 177:105099, 2023.

\bibitem{Harris1963}
T.E. Harris.
\newblock {\em The Theory of Branching Processes}, volume 119 of {\em
  Grundlehren der mathematischen Wissenschaften}.
\newblock Springer Berlin, Heidelberg, 1963.

\bibitem{henk2011mycodiplosis}
D.~Henk, D.~Farr, and M.~Aime.
\newblock \emph{Mycodiplosis} (diptera) infestation of rust fungi is frequent,
  widespread and possibly host specific.
\newblock {\em Fungal Ecology}, 4(4):284–289, 2011.

\bibitem{heydari2010biocontrol}
A.~Heydari and M.~Pessarakli.
\newblock A review on biological control of fungal plant pathogens using
  microbial antagonists.
\newblock {\em Journal of Biological Sciences}, 10(4):273–290, 2010.

\bibitem{jaramillo2013climateimpact}
J.~Jaramillo, E.~Muchugu, F.E. Vega, A.~Davis, C.~Borgemeister, and
  A.~Chabi-Olaye.
\newblock Climate change or urbanization? impacts on a traditional coffee
  production system in east africa over the last 80 years.
\newblock {\em PLOS ONE}, 6(11):e24528, 2011.

\bibitem{jeger2008adaptation}
M.~J. Jeger, P.~J. Wijngaarden, and R.~F. Hoekstra.
\newblock Adaptation to the cost of resistance in a haploid clonally
  reproducing organism: The role of mutation, migration and selection.
\newblock {\em Journal of Theoretical Biology}, 252(4):621–632, 2008.

\bibitem{Kurtz1970}
T.G. Kurtz.
\newblock Solutions of ordinary differential equations as limits of pure jump
  markov processes.
\newblock {\em Journal of Applied Probability}, 7(1):49--58, 1970.

\bibitem{kurtz1972relationships}
T.G. Kurtz.
\newblock Relationships between stochastic and deterministic models for
  chemical reactions.
\newblock {\em Journal of Chemical Physics}, 57(7):2976--2978, 1972.

\bibitem{la1976stability}
J.P. La~Salle.
\newblock {\em The stability of dynamical systems}.
\newblock SIAM, 1976.

\bibitem{madden2007study}
L.V. Madden, G.~Hughes, and F.~Van~den Bosch.
\newblock {\em The Study of Plant Disease Epidemics}.
\newblock American Phytopathological Society (APS Press), 2007.

\bibitem{mailleret2009semi}
L.~Mailleret and V.~Lemesle.
\newblock A note on semi-discrete modelling in the life sciences.
\newblock {\em Philosophical Transactions of the Royal Society A: Mathematical,
  Physical and Engineering Sciences}, 367(1908):4779–4799, 2009.

\bibitem{mccook2019history}
S.~McCook.
\newblock {\em Coffee Is Not Forever: A Global History of the Coffee Leaf
  Rust}.
\newblock Ohio University Press, 2019.

\bibitem{meng2010plantdisease}
X.~Meng and Z.~Li.
\newblock The dynamics of plant disease models with continuous and impulsive
  cultural control strategies.
\newblock {\em Journal of Theoretical Biology}, 266(1):29–40, 2010.

\bibitem{mollison1991dependence}
D.~Mollison.
\newblock Dependence of epidemic and population velocities on basic parameters.
\newblock {\em Mathematical Biosciences}, 107(2):255--287, 1991.

\bibitem{nundloll2010impulsive}
S.~Nundloll, L.~Mailleret, and F.~Grognard.
\newblock Two models of interfering predators in impulsive biological control.
\newblock {\em Journal of Biological Dynamics}, 4(1):102–114, 2010.

\bibitem{Rayner1961}
R.W. Rayner.
\newblock Germination and penetration studies on coffee rust (\emph{Hemileia
  vastatrix} b. \& br.).
\newblock {\em Annals of Applied Biology}, 49(3):497--505, 1961.

\bibitem{salcedo2021elucidating}
S.~Salcedo-Sarmiento, C.E. Aucique-P{\'e}rez, P.R. Silveira, A.A. Colm{\'a}n,
  A.L. Silva, P.S. Corr{\^e}a~Mansur, F.~Rodrigues, Harry~C. Evans, and R.W.
  Barreto.
\newblock Elucidating the interactions between the rust \emph{Hemileia
  vastatrix} and a \emph{Calonectria} mycoparasite and the coffee plant.
\newblock {\em iScience}, 24(4):102352, 2021.

\bibitem{setiawati2021variability}
R.~Setiawati, A.~Widiastuti, A.~Wibowo, and A.~Priyatmojo.
\newblock Variability of \textit{Lecanicillium} spp. mycoparasite of coffee
  leaf rust pathogen (\textit{Hemileia vastatrix}) in indonesia.
\newblock {\em Pakistan Journal of Biological Sciences: PJBS}, 24(5):588–598,
  2021.

\bibitem{singh1984bacillus}
V.~Singh and B.~Deverall.
\newblock \emph{Bacillus subtilis} as a control agent against fungal pathogens
  of citrus fruit.
\newblock {\em Transactions of the British Mycological Society},
  83(3):487–490, 1984.

\bibitem{vanbruggen2016impact}
A.H.C. Van~Bruggen, M.M. He, K.~Shin, V.~Mai, K.C. Jeong, M.R. Finckh, and J.G.
  Morris.
\newblock Environmental and health effects of the herbicide glyphosate.
\newblock {\em Science of The Total Environment}, 616-617:255--268, 2016.

\bibitem{VdDWatmough2002}
P.~{van den Driessche} and J.~Watmough.
\newblock Reproduction numbers and sub-threshold endemic equilibria for
  compartmental models of disease transmission.
\newblock {\em Mathematical Biosciences}, 180(1):29--48, 2002.

\bibitem{vandermeer2009evidence}
J.~Vandermeer, I.~Perfecto, and H.~Liere.
\newblock Evidence for hyperparasitism of coffee rust (\emph{{H}emileia
  vastatrix}) by the entomogenous fungus, {L}ecanicillium lecanii, through a
  complex ecological web.
\newblock {\em Plant Pathology}, 58(4):636--641, 2009.

\bibitem{zambolim2016current}
L.~Zambolim.
\newblock Current status and management of coffee leaf rust in {B}razil.
\newblock {\em Tropical Plant Pathology}, 41(1):1--8, 2016.

\end{thebibliography}

\end{document}